\documentclass[aps,showpacs,showkeys,nofootinbib]{revtex4}
\usepackage{epsfig}
\usepackage[latin1]{inputenc}
\usepackage{float,amsmath,amstext,amsthm,amssymb,amsfonts,slashed}
\usepackage{graphicx}

\begin{document}

\title{Dilepton production from hot hadronic matter in nonequilibrium}

\author{B.~Schenke}
\email{schenke@th.physik.uni-frankfurt.de}

\author{C.~Greiner}
\email{carsten.greiner@th.physik.uni-frankfurt.de}

\affiliation{Institut f\"ur Theoretische Physik, %
   Johann Wolfgang Goethe -- Universit\"at Frankfurt, %
   Max--von--Laue--Stra\ss{}e~1, %
   D--60438 Frankfurt am Main, %
   Germany}

\begin{abstract}

The influence of time dependent medium modifications of low mass
vector mesons on dilepton yields is investigated within a nonequilibrium quantum
field theoretical description on the basis of the Kadanoff-Baym
equations. Time scales for the adaption of the spectral properties
to changing self energies are given and, under use of a model for
the fireball evolution, nonequilibrium dilepton yields from the
decay of $\rho$- and $\omega$-mesons are calculated. In a
comparison of these yields with those from calculations that
assume instantaneous (Markovian) adaption to the changing medium,
quantum mechanical memory effects turn out to be important.

\end{abstract}



\maketitle


\section{Introduction and Motivation}
\label{Intro}

Relativistic heavy ion reactions, as performed at the SIS at GSI,
Darmstadt, the AGS at BNL, the SpS at CERN and the RHIC at BNL,
allow for studying strongly interacting matter under extreme
conditions at high densities and temperatures. One of the main
objectives is the creation and identification of new states of
matter, most notably the quark-gluon plasma (QGP). Photons and
dileptons do not undergo strong final state interactions and thus
carry undistorted information especially on the early hot and
dense phases of the fireball. Photon spectra are a suitable
observable for the temperature of such a system whereas dileptons
have encoded additional dynamic information via their invariant
mass. Particularly in the low mass region, dileptons couple
directly to the light vector mesons and reflect their mass
distribution. They are thus considered the prime observable in
studying mass (de-)generation related to restoration of the
spontaneously broken chiral symmetry. Additionally, vector mesons
are also affected by many-body effects due to coupling to baryonic
resonances. For an overview see \cite{Rapp:1999ej}. Indeed the
CERES experiment at the SPS at CERN \cite{ce95,Agakishiev:1997au}
has found a significant enhancement of lepton pairs for invariant
masses below the pole mass of the $\rho$-meson, giving evidence
for such modifications.

In order to be able to extract precise information from the data,
it is essential to find a thorough mathematical description for
the dilepton production of an evolving fireball of strongly
interacting matter. In particular it is necessary to consider the
fact that especially during the early stages of a heavy ion
reaction the system is out of equilibrium - the medium and hence
the properties of the regarded mesons undergo substantial changes
over time. These scenarios have been described within
Boltzmann-type transport calculations using some quantum
mechanically inspired off-shell propagation in
\cite{Effenberger:1999ay,Effenberger:1999uv,Cassing:1999wx,Cassing:1999mh,Cassing:2000ch,Leupold:1999ga,Leupold:2000ma,Larionov:2001zg}.
In principle, a consistent formulation beyond the standard
quasi-particle approximation is needed - fully ab initio
investigations of off-shell mesonic dilepton production without
any further approximation does not exist so far.

More generally nonequilibrium quantum field theory has become a
major topic of research for describing microscopic transport
processes in various areas of physics (see e.g.
\cite{Juchem:2003bi} and references therein). One major question
deals with how quantum systems eventually thermalize. In
\cite{Juchem:2003bi} the quantum time evolution of $\phi^4$-theory
for homogeneous systems in 2+1 space-time dimensions for far from
equilibrium initial conditions has been investigated, while
earlier works (e.g. \cite{Berges:2000ur}) studied the 1+1
dimensional case. It was shown in \cite{Juchem:2003bi} that the
asymptotic state in the far future corresponds to the exact
off-shell thermal state of the system obeying the equilibrium
Kubo-Martin-Schwinger (KMS) relations among the various two-point
functions. For a coupled fermion-boson Yukawa-type system in 3+1
dimensions, eventual equilibration and thermalization in momentum
occupation was shown in \cite{Berges:2002wr}. In addition, the
full quantum dynamics of the spectral information was analyzed in
\cite{Juchem:2003bi}. This issue was also addressed in some detail
in \cite{Aarts:2001qa}. In a subsequent work \cite{Juchem:2004cs}
the exact solutions were examined in comparison to approximated,
instantaneous off-shell transport equations, obtained by a first
order gradient expansion. In that particular case it turned out
that indeed these approximated equations are a very good
substitute for the full dynamics.

When dealing with vector mesons the important question emerges
whether such a quasi instantaneous adaption of the dynamic and
spectral information to the changing medium, as also assumed in
more schematic model calculations \cite{Rapp:1999us,Rapp:1999ej}
and Monte-Carlo kinetic transport simulations
\cite{Cassing:1999es}, is always a suitable assumption. In a
ultrarelativistic heavy ion collision the typical lifetime of the
diluting hadronic phase is only $5-8$ fm/c \cite{Rapp:1999ej}. On
the other hand the spectral information should always react on
temporal changes with a certain "quantum mechanical" retardation.
If the timescale of these changes of the system becomes comparable
to the retardation time, an instantaneous approximation becomes
invalid and memory effects for the spectral properties of the
excitations are present. Such possible (quantum mechanical) memory
effects, i.e., potential non-Markovian dynamics, have often
appeared in descriptions of the microscopic evolution of complex
quantum systems. Also a nonequilibrium treatment of photon
production from a hot QGP was given in
\cite{Wang:2000pv,Wang:2001xh,Boyanovsky:2003qm,Boyanovsky:2003rw}
and the importance of memory effects and shortcomings of the
S-matrix approach were pointed out for that case. The question of
whether memory effects are important for dilepton production from
hot hadronic matter, i.e., whether such effects have influence on
the measured dilepton yields, constitutes the major motivation for
the present study.

We give for the first time explicit calculations for dilepton
production from first principle nonequilibrium transport
equations. We approach the problem using a nonequilibrium quantum
field theoretical description based on the formalism established
by Schwinger and Keldysh \cite{Schw61,Ke64,Ke65,Cr68}. By this we
set up a framework incorporating the full quantum dynamics, which
is necessary for the description of the transport of true
off-shell excitations. We present a new derivation of the formula
for the dynamic dilepton production rate starting from the
Kadanoff-Baym equations \cite{kb62}, which are nonlocal in time
and hence account for the finite memory of the system. The
resulting formula for the rate involves a (half) Fourier transform
over past times of the two-time Green function of the virtual
photon. The issue within this representation is that the rate for
an invariant mass in the range of interest is hidden as a tiny
component in the two-time function, which contains the rate for
all invariant masses. With this work we meet the challenge of
evaluating this expression for the nonequilibrium dilepton
production rate, which is particularly important since it is the
only causal approach that retains all memory effects. Any
treatment so far, involving gradient expansions of the
Kadanoff-Baym equations, where future contributions to the Green
function are treated equally to those from the past, can not
precisely describe a system that is quickly evolving with respect
to the timescales on that the regarded quantities adjust to system
changes. The reader, who is already familiar with the formalism of
nonequilibrium quantum field theory may skip most of the first
part of Section \ref{noneqrate} and continue reading with
Eq.(\ref{dnlong}).

The paper is organized as follows. We start with a brief
introduction of the used formalism and the presentation of our new
derivation of the dilepton production rate for nonequilibrium
systems in Sections \ref{dileptonproduction} and
\ref{electronself}. The involved two-time Green function of the
virtual photon is further discussed in Section
\ref{inmediumselfenergy}. It is then shown how the medium
modifications of the light vector mesons enter the dilepton rate
via the principle of vector meson dominance. The simulation of
medium modifications of these vector mesons by introduction of a
certain time dependence of their self energies is introduced in
Section \ref{selfenergies}. We analyze the contributions to the
rate in time representation and discuss the precision of the
numerics in Section \ref{timerep}. A quantitative description of
the retardation is given in Section \ref{sec:adaption} by
introduction of time scales, which characterize the memory of the
spectral function. Comparing them to typical time scales in heavy
ion reactions reveals that changes of the spectral function can
not generally be assumed to be adiabatic and that memory effects
can become important. We discuss quantum mechanical interference
effects occurring within this full quantum field theoretical
description in Section \ref{quanint}. Dilepton yields are
calculated first for constant temperature and volume in Section
\ref{sec:consttemp}. Finally, convolution of the dynamic rates
with a fireball model employing a Bjorken like expansion leads to
our most important result, presented in Section \ref{yields}:
Comparison to the quantities computed in the static limit, where
all meson properties adjust to the medium instantaneously, the so
called Markov limit, reveals the significance of memory effects
and the consideration of the full dynamics for certain cases such
as the celebrated and continuously discussed Brown-Rho scaling
\cite{br91}.


\section{The nonequilibrium production rate}
\label{noneqrate}
\subsection{Lepton number transport equation}
\label{dileptonproduction}
    We utilize the Schwinger-Keldysh realtime formalism and the emerging Kadanoff-Baym equations \cite{kb62}
    in order to derive the dynamic nonequilibrium rate of produced
    electron-positron pairs, coming from the decay of light vector
    mesons via virtual photons in a spatially homogeneous, yet time dependent system.
    A different derivation for the dilepton rate
    was performed in \cite{Cooper:1998hu} for dileptons from a pion plasma
    as well as in \cite{Serreau:2003wr}, starting from the dilepton correlator.
    The resulting formulas will provide a
    powerful tool to compute the dynamic behavior of the dilepton
    production rate, influenced by a changing surrounding medium.

    We extract the number of produced
    electrons with momentum $\textbf{p}$ at time $\tau$ from the Wigner transform of the electron
    propagator $G^{<}(1,2)=i\langle\bar{\Psi}(2)\Psi(1)\rangle$ at equal times $t_1=t_2=\tau$, which for a general system is given by (using notation as in \cite{bd65})
\begin{align}
    G^<(\textbf{X},\textbf{p},\tau)&=i\int\frac{d^3q}{(2\pi)^3}\sum_{r s}\frac{m}{\sqrt{E_-E_+}}\left\{\left\langle
    b^{\dag}_{\textbf{p}-\frac{\textbf{q}}{2},r}b_{\textbf{p}+\frac{\textbf{q}}{2},s}\right\rangle u(\textbf{p}+\frac{\textbf{q}}{2},s)\bar{u}(\textbf{p}-\frac{\textbf{q}}{2},r)
    e^{i\textbf{q}\cdot\textbf{X}}e^{i\tau(E_--E_+)}\right.\notag\\
    &\left.+\left\langle b^{\dag}_{\textbf{p}-\frac{\textbf{q}}{2},r}d_{-\textbf{p}-\frac{\textbf{q}}{2},s}\right\rangle
    v(-\textbf{p}-\frac{\textbf{q}}{2},s)\bar{u}(\textbf{p}-\frac{\textbf{q}}{2},r)e^{i\textbf{q}\cdot\textbf{X}}e^{i\tau(E_++E_-)}\right.\notag\\
    &\left.+\left\langle d_{-\textbf{p}+\frac{\textbf{q}}{2},r}d^{\dag}_{-\textbf{p}-\frac{\textbf{q}}{2},s}\right\rangle
    v(-\textbf{p}-\frac{\textbf{q}}{2},s)\bar{v}(-\textbf{p}+\frac{\textbf{q}}{2},r)e^{i\textbf{q}\cdot\textbf{X}}e^{i\tau(E_+-E_-)}\right.\notag\\
    &\left.+\left\langle d_{-\textbf{p}+\frac{\textbf{q}}{2},r}b_{\textbf{p}+\frac{\textbf{q}}{2},s}\right\rangle
    u(\textbf{p}+\frac{\textbf{q}}{2},s)\bar{v}(-\textbf{p}+\frac{\textbf{q}}{2},r)e^{i\textbf{q}\cdot\textbf{X}}e^{i\tau(-E_+-E_-)}\right\}\text{,}
\end{align}
    by projecting on the quantity $\left\langle b^{\dag}_{\textbf{p},r}b_{\textbf{p},s}\right\rangle \delta_{r\,s}$
    \cite{Greiner:1994xm}:
    \begin{equation}
        N(\textbf{p},\tau) = -i \int d^3 X \text{Tr}\left\{\mathcal{P}\,G^{<}(\textbf{X},\textbf{p},\tau)\right\}\text{.}\notag
    \end{equation}
    It can be easily verified that this is achieved by use of the projector
    \begin{align}
        \mathcal{P}=\gamma_0\frac{m}{E_{\textbf{p}}}\sum_{\bar{s}}u(\textbf{p},\bar{s})u^{\dag}(\textbf{p},\bar{s})=\gamma_0\frac{1}{2E_{\textbf{p}}}(\slashed{p}+m)\gamma_0\text{,}\label{projector}
    \end{align}
    where we used $\sum_{\bar{s}}u(\textbf{p},\bar{s})\bar{u}(\textbf{p},\bar{s})=\frac{1}{2m}(\slashed{p}+m)$.
    $E_+=E_{\textbf{p}+\frac{\textbf{q}}{2}}$ and $E_-=E_{\textbf{p}-\frac{\textbf{q}}{2}}$ are the energies corresponding to momentum states
    $\pm(\textbf{p}+\frac{\textbf{q}}{2})$ and $\pm(\textbf{p}-\frac{\textbf{q}}{2})$.
    For further details on the spin decomposition of the Wigner function see \cite{bezz:1972,Botermans:1990qi,Mrowczynski:1992hq}.
    The equations of motion for $G^{<}(1,2)$ and $G^>(1,2)=-i\langle\Psi(1)\bar{\Psi}(2)\rangle$ are the Kadanoff-Baym equations,
    generalized to the relativistic Dirac structure \cite{bezz:1972,Greiner:1994xm,Botermans:1990qi,Mrowczynski:1992hq}:
    \begin{align}
        \left(i\gamma_{\mu}\partial_{1}^{\mu}-m-\Sigma_{HF}(1)\right)G^{\gtrless}(1,1')
        &= \int_{t_0}^{t_1}d2(\Sigma^{>}(1,2)-\Sigma^{<}(1,2))G^{\gtrless}(2,1')\notag\\
        &~~-\int_{t_0}^{t_{1'}}d2\Sigma^{\gtrless}(1,2)(G^{>}(2,1')-G^{<}(2,1'))\text{,}
        \label{transkb1}
    \end{align}
    \begin{align}
        G^{\gtrless}(1,1')\left(-i\gamma_{\mu}\overleftarrow{\partial}_{1'}^{\mu}-m-\Sigma_{HF}(1')\right)
        &= \int_{t_0}^{t_1}d2(G^{>}(1,2)-G^{<}(1,2))\Sigma^{\gtrless}(2,1')\notag\\
        &~~-\int_{t_0}^{t_{1'}}d2 G^{\gtrless}(1,2)(\Sigma^{>}(2,1')-\Sigma^{<}(2,1'))\text{,}
        \label{transkb2}
    \end{align}
    with the self energy $\Sigma$ and its local, Hartree-like term
    $\Sigma_{HF}$. $(1,2)$ is the short term notation for the coordinates $(t_1,\textbf{x}_1,t_2,\textbf{x}_2)$.
    Using retarded and advanced Green functions
    \begin{align}
        G^{\text{ret/adv}}(1,2)=\pm\theta(\pm(t_1-t_2))(G^{>}(1,2)-G^{<}(1,2))\text{,}
    \end{align}
    an important relation can be obtained directly from the Kadanoff-Baym
    equations:
    \begin{align}
        G^{\gtrless}(1,1')=&\int_{t_0}^{\infty}d2\int_{t_0}^{\infty}d3G^{\text{ret}}(1,2)\Sigma^{\gtrless}(2,3)G^{\text{adv}}(3,1')\notag\\
        &+\int d \textbf{x}_2\int d \textbf{x}_3
        G^{\text{ret}}(1,\textbf{x}_2,t_0)G^{\gtrless}(\textbf{x}_2,t_0,\textbf{x}_3,t_0)G^{\text{adv}}(\textbf{x}_3,t_0,1')\text{.}
        \label{fdt}
    \end{align}
    It can be regarded as a generalized fluctuation dissipation relation \cite{bezz:1972,da84,Greiner:1998vd}.
    The second term accounts for the initial conditions at time $t_0$
    only. It can be neglected if one lets the system evolve into a specified initial state for a sufficiently long time. This is done
    by keeping the self energy insertions time independent prior to the onset of the dynamics at the initial time $t_0$.
    Fourier transformation of (\ref{transkb1}) and (\ref{transkb2}) in the spatial coordinates $(\textbf{x}_1-\textbf{x}_{1'})$ and taking
    $\gamma_0(\ref{transkb1})-(\ref{transkb2})\gamma_0$ at $t_1=t_{1'}=\tau$ leads to
    \begin{align}
        i\partial_{\tau} G^{<}(\textbf{p},\tau)& - \textbf{p}\cdot (\gamma_0
        \mbox{\boldmath$\gamma$}G^{<}(\textbf{p},\tau)-G^{<}(\textbf{p},\tau)\mbox{\boldmath$\gamma$}\gamma_0)\notag\\
        & - m(\gamma_0 G^{<}(\textbf{p},\tau)-G^{<}(\textbf{p},\tau)\gamma_0) = \gamma_0 \overrightarrow{C}(\textbf{p},\tau)-\overleftarrow{C}(\textbf{p},\tau)\gamma_0
        \label{transeq1}
        \text{,}
    \end{align}
    with the collision-terms
    \begin{align}
         \overrightarrow{C}(\textbf{p},\tau)&=\int_{t_0}^{\tau}d\bar{t}
         \left(\Sigma^>(\textbf{p},\tau,\bar{t})G^<(\textbf{p},\bar{t},\tau)-\Sigma^<(\textbf{p},\tau,\bar{t})G^>(\textbf{p},\bar{t},\tau)\right)\notag\\
         \overleftarrow{C}(\textbf{p},\tau)&=\int_{t_0}^{\tau}d\bar{t}
         \left(G^>(\textbf{p},\tau,\bar{t})\Sigma^<(\textbf{p},\bar{t},\tau)-G^<(\textbf{p},\tau,\bar{t})\Sigma^>(\textbf{p},\bar{t},\tau)\right)\notag\text{,}
    \end{align}
    where $\Sigma_{HF}$ has been effectively absorbed into the mass $m$.
    Application of the projector (\ref{projector}) to Eq. (\ref{transeq1}) then yields the electron production rate at time $\tau$:
    \begin{align}
        \partial_{\tau} N(\textbf{p},\tau) & = - \text{Tr}\left\{\mathcal{P}(\gamma_0 \overrightarrow{C}(\textbf{p},\tau)-\overleftarrow{C}(\textbf{p},\tau)\gamma_0)\right\}\notag\\
                                           &= (-2) \text{Re} \left[\text{Tr}\left\{\mathcal{P}(\gamma_0 \overrightarrow{C}(\textbf{p},\tau))\right\}\right]\text{,}
        \label{transnumbereqm}
    \end{align}
    Due to having a very long mean free path, the electrons are not expected to interact with the medium after they have been produced.
    This is why they can be described using the free propagators
    \begin{align}
        G^{<}_{0}(\textbf{p},\bar{t},\tau)
        = i \frac{1}{2 E_{\textbf{p}}}(\gamma_0 E_{\textbf{p}} + \mbox{\boldmath$\gamma$}\cdot\textbf{p}-m) e^{i E_{\textbf{p}}(\bar{t}-\tau)}\label{perteq1}\\
        G^{>}_{0}(\textbf{p},\bar{t},\tau)
        = -i \frac{1}{2 E_{\textbf{p}}}(\gamma_0 E_{\textbf{p}} - \mbox{\boldmath$\gamma$}\cdot\textbf{p}+m) e^{-i
        E_{\textbf{p}}(\bar{t}-\tau)}\label{perteq2}\text{.}
    \end{align}
    With that Eq. (\ref{transnumbereqm}) becomes
    \begin{align}
        \partial_{\tau} N(\textbf{p},\tau) &= 2 \text{Re}\left[ \text{Tr}\left\{\mathcal{P}\gamma_0\left(\int_{t_0}^{\tau}d\bar{t}
         \left(\Sigma^<(\textbf{p},\tau,\bar{t})G^{>}_{0}(\textbf{p},\bar{t},\tau)-\Sigma^>(\textbf{p},\tau,\bar{t})G^{<}_{0}(\textbf{p},\bar{t},\tau)\right)\right)\right\}\right]\notag\\
         &=2 \text{Re}\left[ \text{Tr}\left\{\int_{t_0}^{\tau}d\bar{t} \left(\Sigma^<(\textbf{p},\tau,\bar{t})G^{>}_{0}(\textbf{p},\bar{t},\tau)\right)\right\}\right]
         \text{,}\notag
    \end{align}
    where in the second step we used that $G^{>}_{0}(\textbf{p},\bar{t},\tau)\mathcal{P}\gamma_0=G^{>}_{0}(\textbf{p},\bar{t},\tau)$
    and $G^{<}_{0}(\textbf{p},\bar{t},\tau)\mathcal{P}\gamma_0=0$ together with the cyclic invariance of the trace.
    Inserting (\ref{perteq2}) finally leads to
    \begin{align}
        \partial_{\tau} N(\textbf{p},\tau)
        &=2 ~ \text{Im}\left[ \text{Tr}\left\{\frac{\slashed{p}+m}{2 E_{\textbf{p}}}\int_{t_0}^{\tau}d\bar{t} \left(\Sigma^<(\textbf{p},\tau,\bar{t})\right)
        e^{i E_{\textbf{p}}(\tau-\bar{t})}\right\}\right]\text{,}\label{pertfinal}
    \end{align}
    with $p_0 = E_{\textbf{p}}$.
    The full dynamic information for the production of an electron at a given time $\tau$ is incorporated in the memory integral
    from the initial time $t_0$ until the present $\tau$ on the right hand side of Eq. (\ref{pertfinal}).


\subsection{The electron self energy $\Sigma$}
\label{electronself}
\begin{figure}[H]
      \begin{center}
        \includegraphics[width=4cm]{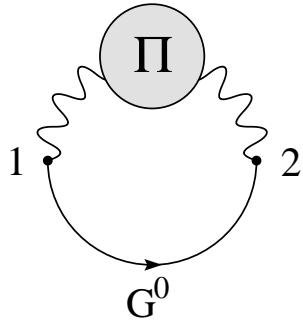}
        \caption{Feynman graph for the electron self energy $\Sigma(1,2)$}
        \label{fig:selfenergy}
      \end{center}
    \end{figure}

    The medium as the source for the production of dileptons enters via the dressing of the virtual photon propagator
    in the electron self energy (see Fig. \ref{fig:selfenergy}). This dressing will finally be given by vector mesons.
    $\Pi$ is the self energy of the virtual photon and we have
    \begin{align}
        i \Sigma^{<}(\textbf{p},t_1,t_2)= - e^2
        \gamma_{\mu}\left(\int
        \frac{d^3k}{(2\pi)^3}D_{\gamma}^{<\,\mu\nu}(\textbf{k},t_1,t_2)G^{<}_{0}(\textbf{p}-\textbf{k},t_1,t_2)\right)\gamma_\nu\text{,}
    \end{align}
    with $D_{\gamma}^{<\,\mu\nu}(\textbf{k},t_1,t_2)$ being the propagator of the dressed virtual photon with momentum $\textbf{k}$. Inserting this
    with the explicit form of the free electron propagator (\ref{perteq1}) into Eq. (\ref{pertfinal}) yields
    \begin{align}
        \partial_{\tau}
        N(\textbf{p},\tau)=2~\text{Re}&\left[e^2\int_{t_0}^{\tau}d\bar{t}\int\frac{d^3k}{(2\pi)^3}(i)D_{\gamma}^{<\,\mu\nu}(\textbf{k},\tau,\bar{t})e^{iE_{\textbf{p}}(\tau-\bar{t})}e^{E_{\textbf{p}-\textbf{k}}(\tau-\bar{t})}\right.\notag\\
        &~~~~\left.\times\frac{1}{2E_{\textbf{p}}}\frac{1}{2E_{\textbf{p}-\textbf{k}}}\text{\text{Tr}}\left\{(\slashed{p}+m)\gamma_{\mu}(\gamma_0E_{\textbf{p}-\textbf{k}}+\mbox{\boldmath$\gamma$}\cdot(\textbf{p}-\textbf{k})-m)\gamma_{\nu}\right\}\right]
        \text{,}
    \end{align}
    where evaluation of the trace leads to
    \begin{align}
        \partial_{\tau}
        N(\textbf{p},\tau)&=2~e^2~\int\frac{d^3k}{(2\pi)^3}\frac{1}{E_{\textbf{p}}}\frac{1}{E_{\textbf{p}-\textbf{k}}}
        \left[p_{\mu}(k-p)_{\nu}+p_{\nu}(k-p)_{\mu}-g_{\mu\nu}(p_{\mu}(k-p)^{\mu}+m^2)\right]\notag\\
        &~~~~\times\text{Re}\left[\int_{t_0}^{\tau}d\bar{t} i
        D_{\gamma}^{<\,\mu\nu}(\textbf{k},\tau,\bar{t})e^{i(E_{\textbf{p}}+E_{\textbf{k}-\textbf{p}})(\tau-\bar{t})}\right]\text{.}
        \label{sigmalessfinal}
    \end{align}
    Defining $p^{-}=k-p$ and $p^{+}=p$ as the four-momenta of the outgoing electron and positron, we rewrite Eq. (\ref{sigmalessfinal}) to
    \begin{align}
        E_+E_-\frac{dR}{d^3p^{+}d^3p^{-}}(\tau)=&\frac{2e^2}{(2\pi)^6}
        \left[p_{\mu}^+p_{\nu}^-+p_{\nu}^+p_{\mu}^--g_{\mu\nu}(p^+p^{-}+m^2)\right]\notag\\
        &~~~~\times\text{Re}\left[\int_{t_0}^{\tau}d\bar{t} i
        D_{\gamma}^{<\,\mu\nu}(\textbf{k},\tau,\bar{t})e^{i(E_+ +
        E_-)(\tau-\bar{t})}\right]\text{,}\label{galekapustarate}
    \end{align}
    with $E_-=E_{\textbf{k}-\textbf{p}}$ and $E_+=E_{\textbf{p}}$.
    $R$ denotes the number of lepton pairs per unit four-volume,
    produced with the specified momentum configuration.
    We now show that Eq. (\ref{galekapustarate}) is indeed the generalization of the well
    known thermal production rate for lepton pairs \cite{McLerran:1984ay,we90,gk91}.
    Using the Fourier transform in relative time coordinates, defined by
    \begin{equation}
        D_{\gamma}^{<\,\mu\nu}(\textbf{k},\tau,\bar{t})
        =\int\frac{d \omega}{2\pi}
        D_{\gamma}^{<\,\mu\nu}(\textbf{k},\tau,\omega) e^{-i \omega
        (\tau -\bar{t})}\text{,}
    \end{equation}
    and taking $t_0 \rightarrow -\infty$, we have for the stationary case:
    \begin{align}
        E_+E_-\frac{dR}{d^3p^{+}d^3p^{-}}=&\frac{2e^2}{(2\pi)^6}
        \left[p_{\mu}^+p_{\nu}^-+p_{\nu}^+p_{\mu}^--g_{\mu\nu}(p^+p^{-}+m^2)\right]\notag\\
        &~~~~\times\text{Re}\left[i \int\frac{d\omega}{2\pi}D_{\gamma}^{<\,\mu\nu}(\textbf{k},\omega)
        \int_{-\infty}^{\tau}d\bar{t} e^{i(E_+ + E_- -\omega)(\tau-\bar{t})}\right]
    \end{align}
    $i D_{\gamma}^{<\,\mu\nu}(\textbf{k},\omega)$ is real and time independent in the stationary case.
    The real part of the last integral is simply $\pi \delta(E_+ + E_- -\omega)$. With that
    and the virtual photon momentum $k^{\mu}=(E,\textbf{k})$, $E=E_++E_-$,
    the rate becomes
    \begin{align}
        E_+E_-\frac{dR}{d^3p^{+}d^3p^{-}}&=-\frac{e^2}{(2\pi)^6}
        \left[p_{\mu}^+p_{\nu}^-+p_{\nu}^+p_{\mu}^--g_{\mu\nu}(p^+p^{-}+m^2)\right] D_{\gamma}^{<\,\mu\nu}(k)\notag\\
        &=-\frac{e^2}{(2\pi)^6}
        \left[p_{\mu}^+p_{\nu}^-+p_{\nu}^+p_{\mu}^--g_{\mu\nu}(p^+p^{-}+m^2)\right]
        \frac{1}{M^4} \Pi_{\gamma}^{<\,\mu\nu}(k)\notag\\
        &=-\frac{2e^2}{(2\pi)^6}
        \left[p_{\mu}^+p_{\nu}^-+p_{\nu}^+p_{\mu}^--g_{\mu\nu}(p^+p^{-}+m^2)\right]\frac{1}{M^4} \frac{1}{e^{\beta E}-1}\text{Im}\Pi_{\gamma}^{\text{ret}\,\mu\nu}(k)\text{,}\label{galekapustarate2}
    \end{align}
    where we used Eq. (\ref{fdt}) for $D_{\gamma}^{<\,\mu\nu}(k)$ in its stationary limit in the first step as well as
    $\Pi^<=2i n_{\text{B}} \text{Im} \Pi^{\text{ret}}$, which follows from the Kubo-Martin-Schwinger (KMS) relation \cite{kubo57,MS59,Greiner:1998vd}, in the second step.
    Eq. (\ref{galekapustarate2}) is the well known rate of dilepton production derived in e.g. \cite{gk91}\footnotemark.
    \footnotetext{The difference in the overall sign is due to the opposite sign in the definition of the Green functions in \cite{gk91}.}

    We return to the nonequilibrium formula (\ref{galekapustarate}) and project on the virtual photon momentum using
    $$\frac{dN}{d^4xd^4k}=\int\frac{dR}{d^3p^+d^3p^-}\delta^4(p^++p^--k)d^3p^+d^3p^-\text{.}$$
    This leads to
    \begin{align}
        \frac{dN}{d^4xd^4k}(\tau,\textbf{k},E)=&\frac{2e^2}{(2\pi)^6}\int\frac{d^3p^+}{E_+}\int\frac{d^3p^-}{E_-}\delta^4(p^++p^--k)
        \left[p_{\mu}^+p_{\nu}^-+p_{\nu}^+p_{\mu}^--g_{\mu\nu}(p^+p^{-}+m^2)\right]\notag\\
        &~~~~\times\text{Re}\left[\int_{t_0}^{\tau}d\bar{t} i
        D_{\gamma}^{<\,\mu\nu}(\textbf{k},\tau,\bar{t})e^{i E(\tau-\bar{t})}\right]\text{}\label{dnlong}
    \end{align}
    for the production rate of dilepton pairs of momentum $k=(E=E_++E_-,\textbf{k})$.
    In the following numerical study we will consider the mode $\textbf{k}=0$ exclusively, i.e., the virtual photon resting with respect
    to the medium. For this case, and taking the electron mass $m$ to zero, Eq. (\ref{dnlong}) can be
    simplified to
    \begin{align}
        \frac{dN}{d^4xd^4k}(\tau,\textbf{k}=0,E)=&\frac{2e^2}{(2\pi)^6}\frac{2}{3}\pi(k_{\mu}k_{\nu}-k^2g_{\mu\nu})
        \text{Re}\left[\int_{t_0}^{\tau}d\bar{t} i D_{\gamma}^{<\,\mu\nu}(\textbf{k}=0,\tau,\bar{t})e^{i E (\tau-\bar{t})}\right]\text{.}
        \label{sigmalessrate}
    \end{align}
    This expression is easily understood. The dynamic information is inherent in the memory
    integral on the right hand side that runs over all virtual photon occupation numbers, Fourier transformed at energy $E$ from the initial time to
    the present. Hence this memory integral determines the full nonequilibrium dilepton production rate at time $\tau$.


\subsection{The in-medium virtual photon self energy $\Pi$}
\label{inmediumselfenergy}
    The dilepton production rate (\ref{sigmalessrate}) involves
    the virtual photon occupation number, expressed by the
    propagator $D_{\gamma}^{<}$. We introduce the dynamic
    medium dependence by dressing this virtual photon propagator
    with the medium dependent $\rho$- or $\omega$-meson. This
    dressing enters via the photon self energy $\Pi^<$ through the fluctuation dissipation relation
    (cf. (\ref{fdt})) for $D_{\gamma}^{<\,\mu\nu}$:
        \begin{align}
            D_{\gamma}^{<\,\mu\nu}&=D_{\gamma}^{\text{ret}\,\mu\alpha}\odot\Pi^<_{\alpha\beta}\odot
            D_{\gamma}^{\text{adv}\,\beta\nu}\text{,}\label{photfdr1a}
        \end{align}
    where $\odot$ implies the integration over intermediate
    space-time coordinates.
    In the medium the vector mesons and virtual
    photons have two possible polarizations relative to their
    momentum in the medium. This leads to two different
    self energies $\Pi_{\text{T}}$ (transverse) and $\Pi_{\text{L}}$ (longitudinal). Introduction of the
    projectors $P_{\text{L}}$ and $P_{\text{T}}$ allows us to split the propagators
    and the self energy into a 3-longitudinal and a
    3-transversal part, relatively to the particle's momentum
    \cite{we90}:
        \begin{align}
            D_{\gamma}^{\text{ret}\,\mu\alpha}=-\frac{P_{\text{T}}^{\mu\alpha}}{k^2-\Pi_{\text{T}}^{\text{ret}}}-\frac{P_{\text{L}}^{\mu\alpha}}{k^2-\Pi_{\text{L}}^{\text{ret}}}\text{~;~~}
            D_{\gamma}^{\text{adv}\,\beta\nu}=-\frac{P_{\text{T}}^{\beta\nu}}{k^2-\Pi_{\text{T}}^{\text{ret}*}}-\frac{P_{\text{L}}^{\beta\nu}}{k^2-\Pi_{\text{L}}^{\text{ret}*}}\label{photdecomp2}
        \end{align}
    and
        \begin{align}
            \Pi^{<}_{\alpha\beta}=-P_{\text{T}\,\alpha\beta}\Pi_{\text{T}}^{<}-P_{\text{L}\,\alpha\beta}\Pi_{\text{L}}^{<}\text{.}
        \end{align}
    The projectors fulfill the usual projector properties $P_{(\text{T}/\text{L})}^2=P_{(\text{T}/\text{L})}$ and $P_\text{T} P_\text{L}=P_\text{L} P_\text{T}=0$.
    With that, Eq. (\ref{photfdr1a}) becomes
        \begin{align}
            D_{\gamma}^{<\,\mu\nu}&=D_{\gamma}^{\text{ret}\,\mu\alpha}\odot(-P_{\text{T}\,\alpha\beta}\Pi_{\text{T}}^{<}-P_{\text{L}\,\alpha\beta}\Pi_{\text{L}}^{<})\odot
            D_{\gamma}^{\text{adv}\,\beta\nu}\notag\\
            &=-D_{\gamma,\text{T}}^{<}P_{\text{T}}^{\mu\nu}-D_{\gamma,\text{L}}^{<}P_{\text{L}}^{\mu\nu}\text{,}\label{photTL}
        \end{align}
    with
       \begin{align}
            D_{\gamma,\text{T/L}}^{<}=\frac{1}{k^2-\Pi_{\text{T/L}}^{\text{ret}}}\odot\Pi_{\text{T/L}}^{<}\odot\frac{1}{k^2-\Pi_{\text{T/L}}^{\text{ret}*}}
            \text{.}\label{photDL}
        \end{align}
    In the case $\textbf{k}=0$, considered later, the longitudinal and transverse
    parts become identical and it follows
        \begin{align}
            D_{\gamma}^{<\,\mu\nu}&=-D^{<}_{\gamma,\text{T}} P_{\text{T}}^{\mu\nu} - D_{\gamma,\text{T}}^{<}
            P_{\text{L}}^{\mu\nu}
            =-D_{\gamma,\text{T}}^{<}(P_{\text{T}}^{\mu\nu}+P_{\text{L}}^{\mu\nu})
            =-D_{\gamma,\text{T}}^{<}\left(g^{\mu\nu}-\frac{k^{\mu}k^{\nu}}{k^2}\right)\text{.}\notag
        \end{align}
    In this case the production rate depends only on the transverse part of the
    virtual photon propagator. The appearing factor of $3$ accounts for the two transverse and one longitudinal
    directions:
        \begin{align}
           \frac{dN}{d^4xd^4k}(\tau,\textbf{k}=0,E)=&\frac{2}{3}\frac{e^2}{(2\pi)^5}(3E^2)
           \text{Re}\left[\int_{t_0}^{\tau}d\bar{t} i D_{\gamma,\text{T}}^{<}(\textbf{k}=0,\tau,\bar{t})e^{i E
           (\tau-\bar{t})}\right]\label{photrate1}
        \end{align}
    For dilepton production $\Pi^{\text{ret}}\propto e^2$ and
    $E$ is the invariant mass of the virtual photon. For the
    cases, we are interested in, it holds
    $|\Pi^{\text{ret}}|\ll E$ and we can approximate
        \begin{equation}
            D_{\gamma,\text{T}}^{<}=D_{\gamma,0}^{\text{ret}}\odot\Pi_{\text{T}}^{<}\odot D_{\gamma,0}^{\text{adv}}\label{conv0}\text{.}
        \end{equation}
    For $\textbf{k} \rightarrow 0$
        \begin{align}
            D_{\gamma,0}^{\text{ret}}(\textbf{k}\rightarrow 0,t)=-\theta(t) t=D_{\gamma,0}^{\text{adv}}(\textbf{k}\rightarrow 0,-t)
        \end{align}
    and we may calculate the virtual photon propagator using the
    transport equation
        \begin{align}
            i
            D^{<}_{\gamma,\text{T}}(\textbf{k}=0,\tau,\bar{t})&=\int_{t_0}^{\tau}dt_1\int_{t_0}^{\bar{t}}dt_2~(\tau-t_1)\left(i\Pi_{\text{T}}^<(\textbf{k}=0,t_1,t_2)\right)(\bar{t}-t_2)\text{.}
            \label{photdgamma}
        \end{align}
    The appearing undamped photon propagators lead to diverging
    contributions from early times, i.e., for low frequencies. In the later numerical calculation these
    contributions turn out to be at least of the order $10^5$ larger than the
    actual (higher frequency) structure. Due to the naturally limited
    numerical accuracy, the higher frequency
    structure would get lost among these early time
    contributions. In order to cure this numerical problem, we
    introduce an additional cutoff $\Lambda$ for the free photon
    propagators, i.e., we perform the replacement:
    \begin{align}
        D_{\gamma,0}^{\text{ret}}(\tau-t_1) = (\tau-t_1) \rightarrow (\tau-t_1) e^{-\Lambda (\tau-t_1)}
        \label{lambdaint}
    \end{align}
    and analogously for $D_{\gamma,0}^{\text{adv}}(t_2-\bar{t})$. In the performed calculations we employ
    $\Lambda\approx 0.3$ GeV.
    The exponential factors lead to a reduction of the rate, which
    we will overcome by renormalizing the final result by multiplication with
 $       \frac{(\omega^2+\Lambda^2)^2}{\omega^4}\text{,}$
    a factor we get from Fourier transforming the convolution
    (\ref{photdgamma}), assuming equilibrium. This will not affect
    the time scales we are interested in, and comparison of the
    dynamically computed rate for the stationary case (constant
    self energy) with the analytic thermal rate shows perfect
    agreement.

    Vector meson dominance (VMD) \cite{sa60,sa69} allows for the
    calculation of $\Pi_{\text{T}}^<$, using the identity between the electromagnetic
    current and the canonical interpolating fields of the vector mesons \cite{klz67}:
        \begin{align}
            J_{\mu}=-\frac{e}{g_{\rho}}m_{\rho}^2\rho_{\mu}-\ldots\text{,}
        \end{align}
    which leads to
        \begin{align}
            \Pi^<_{\alpha\beta}=\frac{e^2}{g_\rho^2}m_{\rho}^4
            D_{\rho\,\alpha\beta}^{<}\text{}\label{photvmd}
        \end{align}
    for the self energy.
    When treating the $\omega$-meson, we use the corresponding self
    energy and propagator.
    We again apply the generalized fluctuation dissipation
    relation (\ref{fdt}) to calculate
        \begin{align}
            D_{\rho,\text{T}}^{<}=D_{\rho,\text{T}}^{\text{ret}}\odot\Sigma^<_{\rho,\text{T}}\odot
            D_{\rho,\text{T}}^{\text{adv}}\text{,}\label{photfdr}
        \end{align}
    with the $\rho$-meson self energy $\Sigma^<_{\rho,\text{T}}$.
    The transverse parts of the retarded and advanced propagators $D_{\rho,\text{T}}^{\text{ret}}(\textbf{k},t_1,t_2)=D_{\rho,\text{T}}^{\text{adv}}(\textbf{k},t_2,t_1)$
    of the vector meson in a spatially homogeneous and isotropic medium follow the equation of motion
        \begin{align}
            \left(-\partial_{t_1}^2-m_{\rho}^2-\textbf{k}^2\right)D_{\rho,\text{T}}^{\text{ret}}(\textbf{k},t_1,t_2)-\int_{t_2}^{t_1}d\bar{t}
            \Sigma^{\text{ret}}_{\rho,\text{T}}(\textbf{k},t_1,\bar{t})D_{\rho,\text{T}}^{\text{ret}}(\textbf{k},\bar{t},t_2)=\delta(t_1-t_2)\text{.}\label{photdgl}
        \end{align}
    In the following we will omit the index T for convenience.

    The dynamic medium evolution is now introduced by hand via a specified time dependent retarded meson self
    energy $\Sigma^{\text{ret}}(\tau, \omega)$ with system time $\tau$ (see Section \ref{selfenergies}). From that the self energy $\Sigma^<$,
    needed for solving Eq. (\ref{photfdr}),
    follows by introduction of an assumed background temperature of the fireball.
    The fireball, constituting the medium, generates the time dependent self energy $\Sigma^{\text{ret}}$ and, assuming a quasi thermalized system,
    the $\rho$-meson current-current correlator $\Sigma^<$ is given via
    \begin{equation}
     \Sigma^<(\tau,\omega,\textbf{k})=2i n_{\text{B}}(T(\tau)) \text{Im} \Sigma^{\text{ret}}(\tau,\omega,\textbf{k})\text{,}
    \end{equation}
    which follows from the KMS relation
     \begin{equation}
        \Sigma^{<}(\omega,\textbf{k})=\mp e^{-\beta\omega}
        \Sigma^{>}(\omega,\textbf{k})\text{,}
    \end{equation}
    being valid for thermal systems \cite{kubo57,MS59,Greiner:1998vd}. $n_{\text{B}}$ is the
    Bose-distribution.
    The assumption of a quasi thermalized background medium is of course rather
    strong, but necessary in order to proceed:
    In principle, for a full nonequilibrium situation the self
    energies $\Sigma^<$ and $\Sigma^>$ for the $\rho$-meson have
    to be obtained self-consistently via e.g. coupling to
    resonance-hole pairs \cite{pe98}, being out of
    equilibrium themselves. (For a realization of true
    nonequilibrium dynamics of a homogeneous system within a $\Phi^4$-theory see
    \cite{Juchem:2003bi} and within a coupled fermion-meson system
    \cite{Berges:2002wr}.)
    For an expanding and inhomogeneous reaction geometry this is still not possible today.
    In any case, explicit calculation of $\Sigma^<$ in the two-time representation will cause even stronger memory
    effects.
    Additionally, in order to simulate a realistic situation by application of a fireball model, a temperature needs to be defined
    and hence local equilibrium has to be assumed.

    The framework, we have established, allows us to
    calculate the dynamic evolution of vector mesons' spectral
    properties, occupation, and the \emph{nonequilibrium} dilepton
    production rate, given an evolving medium with assumed time dependent
    quasi-thermal properties. At this point it is also worthwhile
    to point out that working in the two-time representation has
    great advantages over the mixed representation, in which all
    quantities are expressed by their Wigner transforms. The
    problem in this case is that the Wigner transforms of the
    many appearing convolutions are nontrivial. They can be
    expressed by a gradient expansion to infinite order \cite{Henning:1994qz}.
    Explicit calculations are usually carried out within a first order gradient
    approximation, which is not applicable in our case because the system
    is not evolving slowly with respect to the relevant time
    scales (see Sections \ref{sec:adaption} and \ref{yields} for details on the time scales involved).
    All memory is lost by application of this approximation,
    because when using the mixed representation, full Fourier transformations
    from relative time to its conjugate frequency are involved, which
    imply treating contributions from the future and from the past on
    equal grounds. This violates causality in a quickly evolving system.
    Hence, calculations within the two-time representation allow
    for the most exact investigation of the evolving system under
    consideration.


\section{Nonequilibrium dilepton production from an evolving medium}
\label{evmedium}
\subsection{Vector meson self energies}
\label{selfenergies} We are now able to calculate the evolution of
the spectral properties of vector mesons in a changing medium as
well as the corresponding dilepton production rates. We will treat
mass shifts described by Brown-Rho scaling \cite{br91},
broadening, as caused by pion scattering, and scattering of the
mesons with nucleons, leading to further broadening and excitation
of resonances \cite{hf92,cs92,ra96,fp97,kl97,pe98,po04}. The
medium effects are introduced via a specific time evolving self
energy for the vector mesons. The main purpose of this work is to
investigate medium modifications dynamically for the first time,
and compare the results to those obtained from instantaneous,
Markovian calculations.

In the following we first employ simplified self energies for possible broadening
such as
        \begin{align}
            \text{Im}\Sigma^{\text{ret}}(\tau,\omega)=-\omega\Gamma(\tau)\text{,}\label{bwself}
        \end{align}
with a $\textbf{k}$- and $\omega$-independent width $\Gamma$, which, in the static case,
leads to a Breit-Wigner distribution of the spectral function
        \begin{align}
            A(\omega,\textbf{k})&=-\frac{1}{\pi}\text{Im}D_{\rho}^{\text{ret}}(\omega,\textbf{k})=
            \frac{1}{\pi}\frac{\omega \Gamma}
            {(\omega^2-\textbf{k}^2-m^2)^2+(\omega \Gamma)^2}\text{.}\label{modelsbreitwigner}
        \end{align}
The self energy (\ref{bwself}) is given in a mixed time-frequency
representation. The time dependence is being accounted for by
introduction of the system time $\tau$, for which we will discuss
two possible choices below. $\omega$ stems from the Fourier
transformation in relative time $(t_1-t_2)$.

Modifications of the mass are introduced directly as via a local
term in the self energy. From this sort of self energy several
numerical issues emerge. In Eq. (\ref{photdgl}) the self energy
enters in time representation
        \begin{equation}
            \Sigma^{\text{ret}}(t_1,t_2)=(\Sigma^>(t_1,t_2)-\Sigma^<(t_1,t_2))\theta(t_1-t_2)\text{,}
        \end{equation}
Its Fourier transform in relative time $(t_1-t_2)$, $\Sigma^{ret}(\tau,\omega)$, is given by a convolution of
the $\theta$-function's Fourier transform and $(\Sigma^>-\Sigma^<)(\tau,\omega)$:
        \begin{align}
            \Sigma^{\text{ret}}(\tau,\omega)&=i\int\frac{d\bar{\omega}}{2\pi}\frac{1}{\omega-\bar{\omega}+i\epsilon}(\Sigma^>-\Sigma^<)(\tau,\bar{\omega})\notag\\
                                &=i\mathcal{P}\int\frac{d\bar{\omega}}{2\pi}\frac{1}{\omega-\bar{\omega}}(\Sigma^>-\Sigma^<)(\tau,\bar{\omega})
                                +
                                \frac{1}{2}(\Sigma^>-\Sigma^<)(\tau,\omega)\text{,}
        \end{align}
where we used the standard relation
        \begin{equation}
            \frac{1}{\omega-\bar{\omega}+i\epsilon}=\mathcal{P}\frac{1}{\omega-\bar{\omega}}-i
            \pi
            \delta(\omega-\bar{\omega})\text{.}
        \end{equation}
With $(\Sigma^>-\Sigma^<)(\tau,\omega)=-2i\omega\Gamma(\tau)$, one
has
        \begin{align}
           \Sigma^{\text{ret}}(\tau,\omega)&=2\mathcal{P}\int_{-c}^{+c}\frac{d\bar{\omega}}{2\pi}
            \frac{\bar{\omega}}{\omega-\bar{\omega}}\Gamma(\tau)-i\omega\Gamma(\tau)\notag\text{,}
        \end{align}
which has the imaginary part (\ref{bwself}). In addition we also
get an "unwanted" dispersive real part that causes an (infinite)
mass shift. The introduced cutoff $c$ cures that infinity and we
can renormalize the mass by the replacement
        \begin{equation}
            m\rightarrow\sqrt{m^2-\text{Re}\Sigma^{ret}(\omega,\tau)}\text{,}\notag
        \end{equation}
with
        \begin{align}
            \text{Re}\Sigma^{\text{ret}}(\tau,\omega)&=\text{Re}\left[2\Gamma(\tau)\mathcal{P}\int_{-c}^{+c}\frac{d\bar{\omega}}{2\pi}\frac{\bar{\omega}}{\omega-\bar{\omega}}\right]\notag\\
            &=-\frac{2}{\pi}\Gamma(\tau) c
            \left(_2F_1\left(-\frac{1}{2},1,\frac{1}{2},\frac{\omega^2}{c^2}\right)\right)\notag\\
            &\approx-\frac{2}{\pi}\Gamma(\tau) c
            \left(1-\frac{\omega^2}{c^2}-\frac{\omega^4}{3c^4}\right)\text{,}
        \end{align}
with the hypergeometric function $_2F_1$, Taylor expanded to second order in the last line.
To make the renormalization $\omega$-independent, we replace $\omega$ by the physical pole mass $m$, causing only minimal
inaccuracies far from the peak position.

Another potential problem is the behavior of the self energy
$\Sigma^<(\tau,\omega)$ for negative frequencies. As already
described, for a thermalized system, it is given by
$\Sigma^<(\tau,\omega)=2i n_{\text{B}}(T(\tau),\omega) \text{Im}
\Sigma^{\text{ret}}(\tau,\omega)$. For negative frequencies the
propagator $D^<(\tau,\omega)$ contains a factor of $(1+n(\omega))$
since there holds the symmetry relation
$D^<(\tau,-\omega)=D^>(\tau,\omega)$ for the scalar boson case and
$D^>(\tau,\omega)=2 i (1+n_{\text{B}}) A(\tau,\omega)$ in
equilibrium. The first term in parentheses leads to divergent
vacuum contributions, which are uninteresting for the
investigation of the positive frequency behavior and cause
numerical problems. To see this, we first give the limits of the
expression $\Sigma^<(\tau,\omega)$ in frequency representation:
    \begin{align}
        i \Sigma^<(\tau,\omega) = \left\{ \begin{array}{rl} \propto e^{-\omega/T(\tau)} & \text{for } \omega \gg 0 \\
                                                    2 T(\tau) \Gamma(\tau) & \text{for } \omega=0 \\
                                                    -2 \omega \Gamma(\tau) & \text{for } \omega \ll 0 \end{array}
                                                    \right.
    \end{align}
The Fourier transform thus becomes a $\delta'$-distribution for the negative frequency part.
In order to cure this problem we first split the self energy into positive and negative frequency parts
    \begin{align}
        i \Sigma^<(\tau,\omega) &=
        \underbrace{\frac{2 |\omega|
        \Gamma(\tau)}{e^{|\omega|/T(\tau)}-1}}_{=:i\Sigma^{<,\text{eff}}(\tau,\omega)}-2\omega\Gamma(\tau)\theta(-\omega)\text{,}\label{inmediumsigmaeff}
    \end{align}
and neglect the negative frequency part, violating the mentioned
symmetry relation for $D^<$ and $D^>$ (also for $\Sigma^<$ and
$\Sigma^>$). This violation however only means an omittance of the
vacuum contributions for negative frequencies and does not cause
changes for positive frequencies. The same approach may be taken
with other types of self energies having the same symmetry
property under the exchange $\omega\leftrightarrow -\omega$.

\begin{figure}[htb]
      \begin{center}
        \includegraphics[height=7cm]{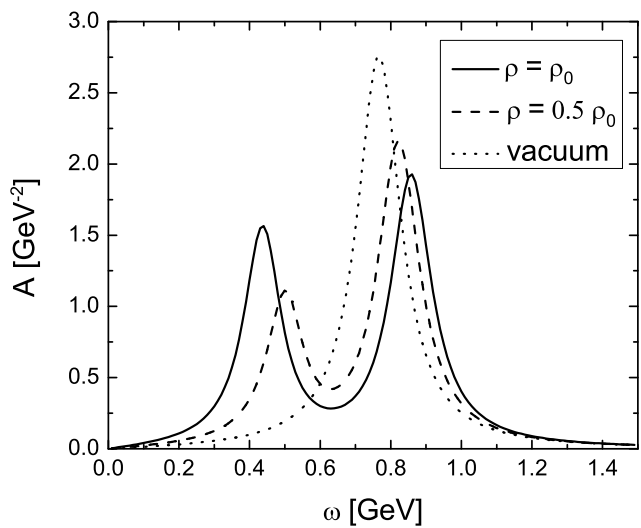}
        \caption{Spectral function of the $\rho$-meson coupled to an N(1520)-hole pair for different densities. $\bar{E}=582$ MeV,$\Gamma_R=120$ MeV}
        \label{fig:markovAresonanceC}
      \end{center}
    \end{figure}

Coupling of resonances to the $\rho-N$-channel  has been treated
by \cite{fp97,pe98,po04}. For the $\textbf{k}=0$ mode, the full
self energy for coupling to $J^P=\frac{3}{2}^-$ -resonances is
given by \cite{po04} (also see \cite{he94}, who treated coupling
of pions to resonance hole excitations)
        \begin{align}
            \text{Im}\Sigma(\tau,\omega,\textbf{k}=0)=-\frac{\rho(\tau)}{3}
            \left(\frac{f_{RN\rho}}{m_{\rho}}\right)^2 g_I
            \frac{\omega^3 \bar{E}\Gamma_R(\tau)}{(\omega^2-\bar{E}^2-\frac{\Gamma_R(\tau)^2}{4})^2+(\Gamma_R(\tau)\omega)^2}-\omega\Gamma(\tau)
            \label{modelshenself}\text{.}
        \end{align}
$\bar{E}=\sqrt{m_R^2+\textbf{k}^2}-m_N$ is the energy necessary
for the meson to scatter from a nucleon at rest, with $m_R$ and
$m_N$ the masses of the resonance and the nucleon respectively.
$\Gamma_R$ is the width of the resonance and $g_I$ is the isospin
factor ($2$ for isospin $\frac{1}{2}$ and $\frac{4}{3}$ for
isospin $\frac{3}{2}$ resonances). $\rho(\tau)$ is the time
dependent density of the system and appears due to application of
the $\rho$-$\cal T$-approximation, where $\cal T$ stands for the
forward scattering amplitude. The important part for our purpose
is the structure of the denominator, which represents a
characteristic pole structure. We will replace $\omega^2$ in the
numerator by a constant factor since it causes divergence of
$\text{Re}\Sigma^{\text{ret}}$, only cured by application of
subtracted dispersion relations corresponding to counterterms. The
cutoff $c$ also prevents divergences but leaving in the factor of
$\omega^2$ causes inaccurate numerical results. We show the
spectral function for the case of the $N(1520)$ resonance, where
$f_{RN\rho}\approx 7.0$ \cite{pe98} and the width of the resonance
is $\Gamma_R=120$ GeV, in Fig. \ref{fig:markovAresonanceC} for
different densities at $\textbf{k}=0$.

We now discuss the system time $\tau$. Defining
$\tau=\frac{t_1+t_2}{2}$, as first done in \cite{kb62}, seems to
be a sensible choice, but when investigating the spectral
function, one is dealing with retarded quantities and we do not
want them to collect information from the future, i.e., we want to
retain causality. This is why we choose $\tau=t_1$ instead of the
symmetric form $\tau=\frac{t_1+t_2}{2}$, because in this case at a
certain time $\tau$ information of $\Sigma^{\text{ret}}$ that is
located in the future of $\tau$ enters the spectral function. This
cannot happen with $\tau=t_1$ as demonstrated in Appendix
\ref{app:acausal}.

For the self energy $\Sigma^<(\tau,\omega)$ one should stick to
the symmetric choice in order to fulfill the symmetry relation
$\Sigma^<(t_1,t_2)=\Sigma^>(t_2,t_1)$. However, doing this reduces
accuracy in the numerical calculation, due to additional necessary
Fourier transformations. Comparison of calculations using either
$\Sigma^<(t_1,\omega)$ or $\Sigma^<(\frac{t_1+t_2}{2},\omega)$
reveals minor differences in the rate when changing the
temperature with system time $t_1$ or $\frac{t_1+t_2}{2}$
respectively, but on average, and hence in the final yield, the
differences cancel, as could be expected.

\subsection{Contributions to the rate in time representation}
\label{timerep} 
\begin{figure}[htb]
      \begin{center}
        \includegraphics[height=6cm]{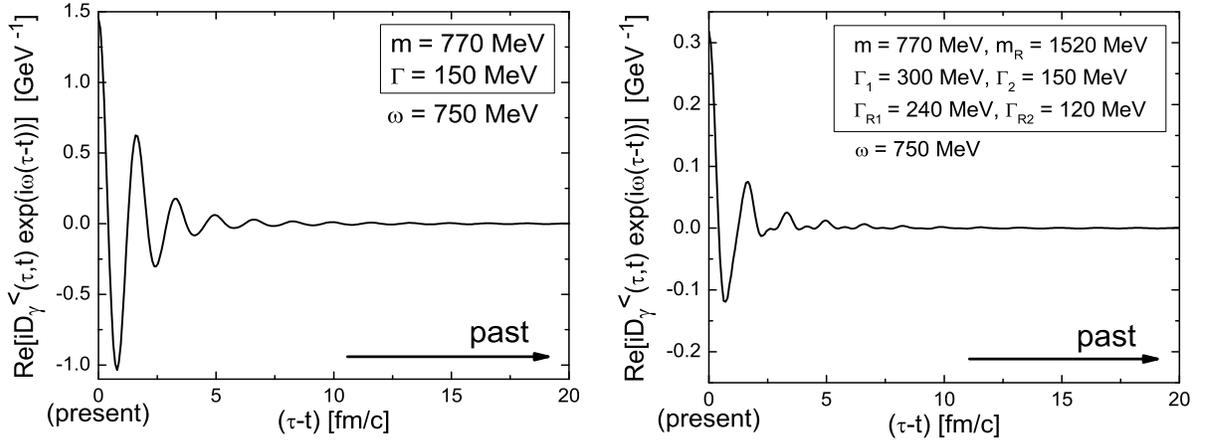}
        \caption{Contributions to the rate for a fixed frequency at given time $\tau$ from times $(\tau-t)$ in the past for the case of
        free $\rho$-mesons at constant temperature (left)
        and the case where the coupling to the N(1520) was turned off 10 fm/c before the present time (widths were also changed as indicated
        in the figure (see text) - index 1 refers to initial, 2 to final
        quantities) (right).}
        \label{fig:intcontconst}
      \end{center}
    \end{figure}

\begin{figure}[htb]
      \begin{center}
        \includegraphics[height=5.8cm]{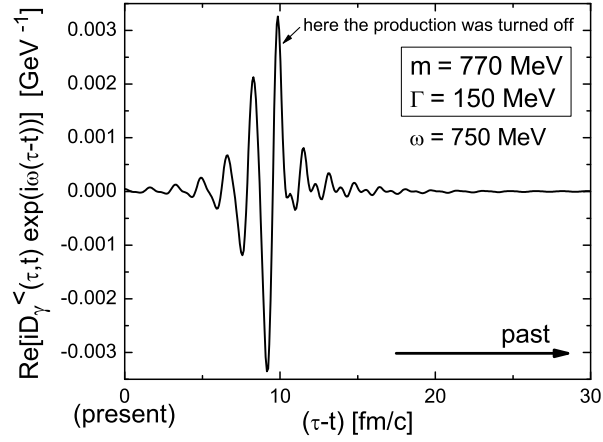}
        \caption{Contributions to the rate for a fixed frequency at given time $\tau$ from times $(\tau-t)$ in the
        past. The production has been turned off 10 fm/c before the present time $\tau$.}
        \label{fig:intcontconstoff}
      \end{center}
    \end{figure}

\begin{figure}[htb]
      \begin{center}
        \includegraphics[height=6cm]{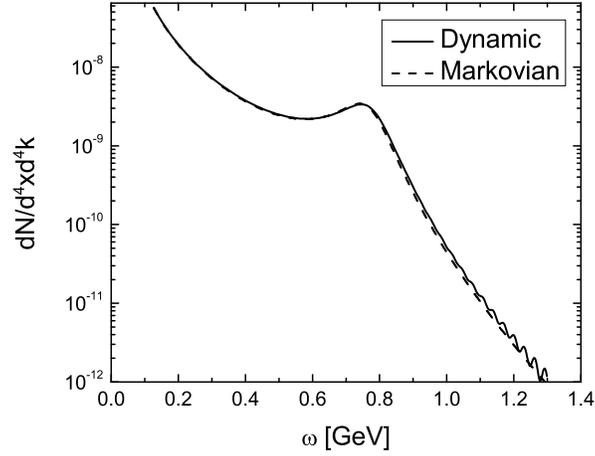}
        \caption{Comparison of the dynamically calculated rate (solid) to the one calculated from Eq. (\ref{markovrate}), the Markovian rate
        (dashed), shows very good agreement.}
        \label{fig:ratenumerics}
      \end{center}
    \end{figure}

Before we calculate nonequilibrium rates and
yields, we investigate how the present rate is created over time,
i.e., we focus on which contributions to Eq. (\ref{photrate1})
come from which times in the past. Figs. \ref{fig:intcontconst}
and \ref{fig:intcontconstoff} show the integrand for fixed energy
$\omega=750$ MeV and fixed time $\tau$ for relative times
$\tau-t$. The first case shown (left of Fig.
\ref{fig:intcontconst}) is an equilibrium scenario with free
$\rho$-mesons embedded in an environment at constant temperature.
One can see that the main contribution comes from the very near
past, but that there are also contributions from early times as
well as large oscillations leading to alternating positive and
negative contributions. The second case (right of Fig.
\ref{fig:intcontconst}) represents $\rho$-mesons, embedded in a
constant heat bath, broadened and coupled to the N(1520) resonance
at the time where the coupling and broadening was completely
turned off again 10 fm/c ago in a way that will be discussed in
Section \ref{sec:adaption} (over a time of $\Delta\tau=7.18$ fm/c,
cf. Fig. \ref{fig:linearchange} - We use this duration because it
results as the lifetime of the hadronic phase of the fireball from
calculations described in Section \ref{yields} and will hence be
used in calculations of the dilepton yields). The integrand shows
richer structures than in the equilibrium case, caused by the
changing self energy in the past, which is quite remarkable: It
shows that after having exposed the $\rho$-mesons to vacuum
conditions for 10 fm/c, a memory of the situation in the more
distant past is remaining. The third case (Fig.
\ref{fig:intcontconstoff}) again represents free $\rho$-mesons but
with the production in the heat bath, caused by $\Sigma^<_{\rho}$,
turned off $10$ fm/c in the past. Comparison of the overall
amplitude to that in Fig. \ref{fig:intcontconst} (left) shows that
the rate at time $\tau$, resulting from the integral over the
function shown in Fig. \ref{fig:intcontconstoff} (see Eq.
(\ref{photrate1})), will be very small compared to the rate before
production was turned off, which is clear since the $\rho$-meson
has a decay constant of 1.3 fm/c with that the rate is
exponentially suppressed after production is turned off. The shape
of the contribution is very interesting: Between the point where
production was turned off and the present the integrand has a
decaying (and oscillating) shape and contributions from the more
distant past also remain. However, the largest values,
contributing to the present rate, are situated close to the time,
at which production was turned off. This is intuitively clear. The
dileptons produced at the present time $\tau$ most dominantly
originate from decaying $\rho$-mesons, produced slightly before
the time when production was turned off. Thereby the strong memory
to that past time is visible in the integrand.
  \begin{figure}[H]
\end{figure}
A full interpretation of the various structures shown is difficult
and it becomes clear that only the completely integrated yield
represents a physical quantity. There is a dependence on the
parameter $\Lambda$ (cf. Eq. (\ref{lambdaint}) and related
discussion) such that contributions from times further in the past
are reduced using larger cutoffs $\Lambda$. On the other hand the
accuracy of the numerics is strongly increased by employing larger
cutoffs (when setting $\Lambda$ to zero all the principal
information is destroyed by diverging contributions, several
orders of magnitude larger than the actual structure responsible
for the creation of the rate). The early time contributions
translate to diverging contributions for the smallest frequencies
after Fourier transformation. The comparison of the rate from a
free $\rho$-meson at constant temperature (i.e., the equilibrium
case) calculated dynamically and the one calculated using the
equilibrium formula
\begin{equation}
    \frac{dN}{d^4x d^4k}(\omega)=\frac{2 e^4}{(2\pi)^5}\frac{m_{\rho}^4}{g_{\rho}^2}\frac{1}{\omega^2}n_{\text{B}}(\omega)\pi A(\omega)\text{,}
    \label{markovrate}
\end{equation}
shows that the dynamically calculated rates, being integrals over strongly oscillating functions, as shown in Figs. \ref{fig:intcontconst} and \ref{fig:intcontconstoff},
are reproduced very well by the numerics as nicely seen in Fig. \ref{fig:ratenumerics}. Here one also notices the large contributions at low frequencies,
which cause the mentioned numerical uncertainties and therefore have to be reduced by introduction of the mentioned cutoff $\Lambda$.
 These contributions are being many orders of magnitude larger than those for frequencies at
the upper end of the regarded range.

\subsection{Time scales of adaption for the spectral function, occupation number and dilepton rate}
\label{sec:adaption}
\begin{figure}[htb]
      \begin{center}
        \includegraphics[height=6.5cm]{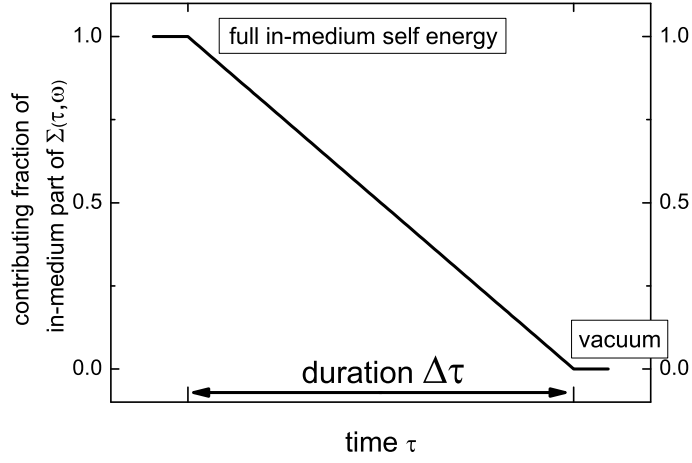}
        \caption{Linear switching off of in-medium effects over a certain duration $\Delta \tau$}
        \label{fig:linearchange}
      \end{center}
    \end{figure}

In order to quantify the times that the mesons' spectral
properties need to adjust to the evolving medium, we regard the
cases of broadening and mass shifts and introduce a time dependent
self energy that represents linear changes (see Fig.
\ref{fig:linearchange}) in width or mass.
We do not perform sudden changes using step functions in time
because that would cause strong, hard to control oscillations in
the Fourier transforms. Apart from the linear function, totally
smooth functions like a hyperbolic tangent have been tested for
the representation of the time evolution. It was found that the
two kinks in the linear function do not cause additional problems
when Fourier transforming. Hence, we chose this representation for
the time evolution because it allows for the precise definition of
the duration of the change, and by that of the time scale, which
is not possible for absolutely smooth functions.

The spectral function is given by the imaginary part of the
Fourier transform in relative time coordinates of the retarded
meson-propagator
\begin{equation}
    D_{\rho}^{\text{ret}}(t_1,t_2)\rightarrow D_{\rho}^{\text{ret}}(\tau,t_1-t_2)\rightarrow-\frac{1}{\pi}\text{Im}D_{\rho}^{\text{ret}}(\tau,\omega)=A(\tau,\omega)\text{,}
\end{equation}
where the system time $\tau$ is chosen to be $t_1$ for reasons
discussed at the end of Section \ref{selfenergies} and in Appendix
\ref{app:acausal}. Taking $\tau$ to be $\frac{t_1+t_2}{2}$ leads
to an about two times faster adjustment of the spectral function,
which is caused by the fact that it is now able to gain
information on the medium symmetrically from the past and the
future, which is unphysical.

\begin{figure}[htb]
      \begin{center}
        \includegraphics[width=9cm]{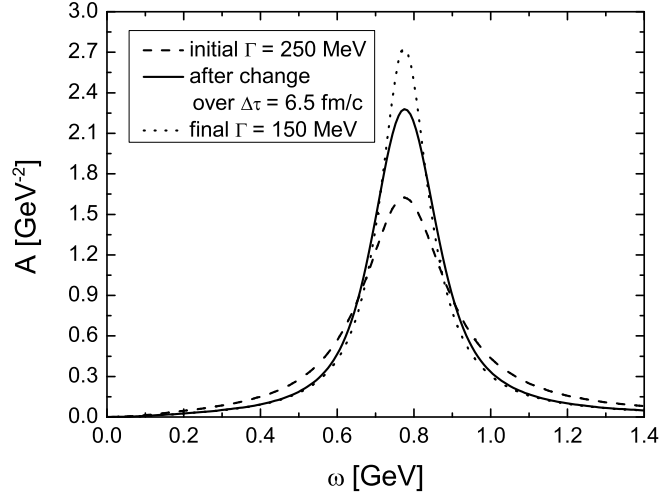}
        \caption{The two compared spectral functions right after the full change of the self energy is reached: dynamic (full) and Markovian (equals the final
        spectral function with $\Gamma=150$ MeV) (dotted) calculation.
         The initial spectral function (dashed) is also shown.}
        \label{fig:width}
      \end{center}
    \end{figure}

The spectral functions's retardation with respect to medium
changes is found by comparing the dynamically calculated spectral
function at the time $\tau_{\text{off}}$ when the medium effects
are fully turned off and the one calculated at the same time,
assuming an instantaneous adaption. This spectral function is
given directly by
        \begin{align}
            A_{\text{Markov}}(\tau,\omega)&=-\frac{1}{\pi}\text{Im}D_{\rho\,\text{Markov}}^{\text{ret}}(\tau,\omega)
                                      =\frac{1}{\pi}\frac{-\text{Im}\Sigma_{\rho}^{\text{ret}}(\tau,\omega)}{(\omega^2
                                      -m_{\rho}^2-\text{Re}\Sigma_{\rho}^{\text{ret}}(\tau,\omega))^2+(\text{Im}\Sigma_{\rho}^{\text{ret}}(\tau,\omega))^2}
        \end{align}
The difference can be made explicit by calculating the difference
in the moment $\int_0^{\infty} A(\tau_{\text{off}},\omega)^2
\omega^2 d\omega$ of the two spectral functions or the difference
in the peak position or height for mass shifts or broadening
respectively. As an illustration of the procedure the spectral
functions, which are compared, are shown for a particular example
of in-medium broadening in Fig. \ref{fig:width} (full and dotted
line). All methods lead to similar results. We find an
exponentially decreasing difference with increasing duration of
the change $\Delta\tau$ (see Fig. \ref{fig:linearchange}). From
this exponential drop we extract a time constant $\bar{\tau}$,
that for the case of broadening (and constant masses) is shown in
Table \ref{tab:scales} for different scenarios. Table
\ref{tab:scalemass} shows time scales for the adjustment of the
spectral function to changes in the meson's mass, extracted from
the peak position. Due to oscillations in the changing spectral
function (see below) an exact extraction of a time scale is more
complicated in this case, but the numbers still give a magnitude
for the retardation.

In the case of the $\rho$-meson
($\Gamma_{\text{vac}}\approx\Gamma_2=150$ MeV) we find a time
scale of about 3 fm/c. That means that the behavior of the $\rho$
mesons becomes adiabatic only for medium changes that are slow
compared to the time of 3 fm/c, i.e., the spectral properties
follow the changes in the medium nearly instantaneously only if
the evolution is very slow as compared to the derived time scale.
For real heavy ion reactions this means that the spectral
properties of vector mesons will retain a certain memory of the
past, and even if they decay outside the medium, they still carry
a certain amount of information on the medium in that they were
produced. This can become important, especially for
$\omega$-mesons, having a vacuum width of $8.49$ MeV \cite{pdb}.
Nevertheless, due to oscillations in the spectral function and
rate and occurring interferences when calculating the dilepton
yield (see below), this strong retardation may be put into
perspective because the only physically meaningful influence of
this retardation is that on the resulting yield (see Section
\ref{yields}). We find the relation
\begin{equation}
 \bar{\tau} \propto \frac{c}{\Gamma_2} \text{, with $c$ between $2$ and $3.5$.}\label{timescale}
\end{equation}

    \begin{table}[htb]
      \begin{center}
        \begin{tabular}{|c|c|c|c|c|c|c|}  \hline
          $\Gamma_2 [\text{MeV}]$ & $\bar{\tau}$ [fm/c] for      & $\bar{\tau}$ [fm/c] for      & $\bar{\tau}$ [fm/c] for     & $\bar{\tau}$ [fm/c] for      & $\bar{\tau}$ [fm/c] for      & $\bar{\tau}$ [fm/c] for      \\
                           & $m=570\text{ MeV}$        & $m=570\text{ MeV}$        & $m=570\text{ MeV}$       & $m=770\text{ MeV}$        & $m=770\text{ MeV}$        & $m=770\text{ MeV}$        \\
                           & $\Gamma_1=200\text{ MeV}$ & $\Gamma_1=300\text{ MeV}$ & $\Gamma_1=400\text{ MeV}$       & $\Gamma_1=200\text{ MeV}$ & $\Gamma_1=300\text{ MeV}$ & $\Gamma_1=400\text{ MeV}$
        \\ \hline
         150 & 2.4 & 3   & 3    & 2.8 & 3.6 & 3.2 \\
         90  & 5   & 5.2 & 6    & 4.8 & 5.4 & 6.2 \\
         70  & 6.4 & 7.2 & 8.2   & 6.4 & 7.2 & 8.2 \\
         50  & 9.4 & 10.8& 12.8 & 8.8 & 10.4& 12.2\\
         30  & 16.4& 20.4& 24.8 & 13.6& 17.4& 21.2
        \\ \hline
        \end{tabular}
      \end{center}
      \caption{Values of $\bar{\tau
      }$ for different masses, initial widths $\Gamma_1$ and final widths $\Gamma_2$.} \label{tab:scales}
    \end{table}

    \begin{table}[htb]
      \begin{center}
        \begin{tabular}{|c|c|c|c|}  \hline
          $\Delta m [\text{MeV}]$ & $\bar{\tau}$ [fm/c] for      & $\bar{\tau}$ [fm/c] for      & $\bar{\tau}$ [fm/c] for      \\
                           & $\Gamma=100\text{ MeV}$ & $\Gamma=150\text{ MeV}$ & $\Gamma=300\text{ MeV}$
        \\ \hline
         100 & 4.33& 3.12 & 2.15  \\
         300 & 5.5 & 3.82 & 2.14  \\
         400 & 6.5 & 3.97 & 2.16
        \\ \hline
        \end{tabular}
      \end{center}
      \caption{Values of $\bar{\tau}$ for different widths and mass shifts $\Delta m$ to masses below the final vacuum mass $m_2=770$ MeV.} \label{tab:scalemass}
    \end{table}

$c$ depends on the other quantities $m$ and $\Gamma_1$ such that
for a larger change in width, the spectral function takes longer
to adjust to it. The reason for the time scale to be (at least)
$2/\Gamma$ is that the spectral function is given by the imaginary
part of the retarded propagator, which, representing a probability
amplitude, is proportional to $e^{-\frac{1}{2}\Gamma t}$. Its
square, an actual probability, is proportional to $e^{-\Gamma t}$,
giving the appropriate decay rate.

\begin{figure}[htb]
      \begin{center}
        \includegraphics[width=8cm]{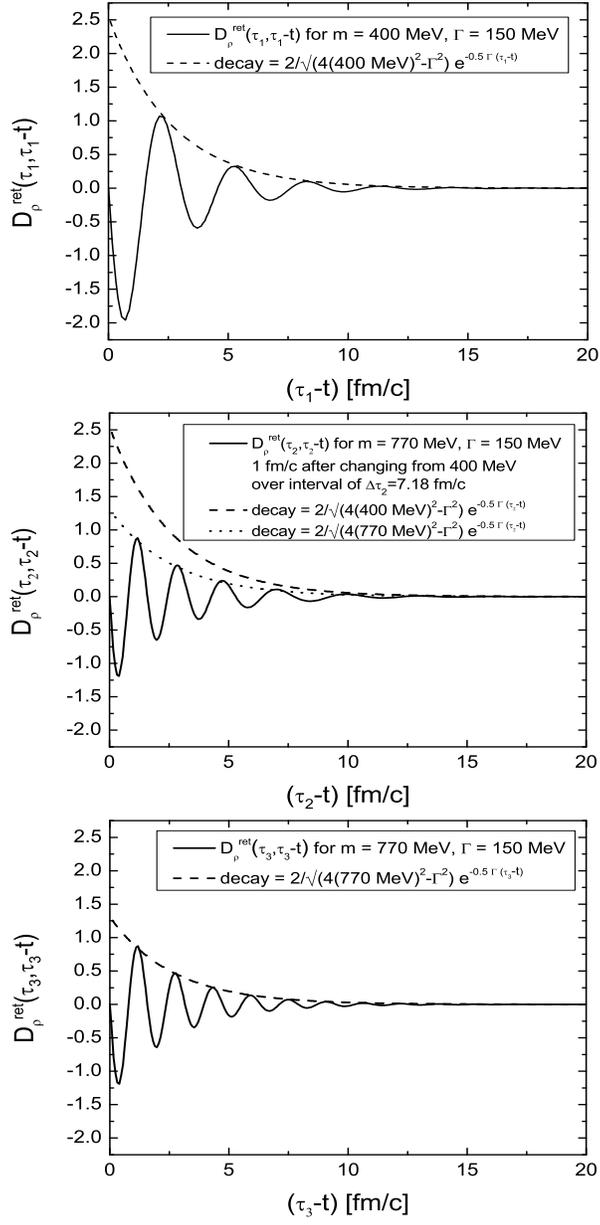}
        \caption{Numerical results for the retarded meson propagator in the two-time representation for a constant mass of 400 MeV, shortly (1 fm/c) after a mass shift to 770
        MeV, and a rather long time (about 75 fm/c) after that mass shift. Discussion in the text.}
        \label{fig:gretdecay}
      \end{center}
    \end{figure}

This result can also be easily retrieved analytically when we
assume a Breit-Wigner type self energy with constant width as well
as a constant mass and solve the equation of motion that follows
directly from Eq. (\ref{photdgl})
\begin{equation}
   \left(-\partial_{t_1}^2-m_{\rho}^2\right)D_{\rho,\text{T}}^{\text{ret}}(t_1,t_2)-\partial_{t_1}D_{\rho,\text{T}}^{\text{ret}}(t_1,t_2)\Gamma=\delta(t_1-t_2)
\end{equation}
by
\begin{equation}
    D_{\rho,\text{T}}^{\text{ret}}(t_1,t_2)=-2\Theta(t_1-t_2)\frac{1}{\sqrt{4 m^2-\Gamma^2}}e^{-\frac{1}{2}\Gamma(t_1-t_2)} \text{sin}(\frac{1}{2}\sqrt{4m^2-\Gamma^2}(t_1-t_2)) \label{dglsol}
\end{equation}
and we find the decay width to be $\Gamma/2$. The numerical
solution shows the same behavior, but here we can go further and
apply changes e.g. to the mass and see how the propagator behaves
in time. In Fig. \ref{fig:gretdecay} we show the numerical
solution for the retarded propagator in time representation
$D_{\rho,T}^{\text{ret}}(\tau,t)$ at different fixed times $\tau$,
for a constant width $\Gamma=150$ MeV and a mass, changed from
$400$ to 770 MeV ($\tau_1$ before the change of duration $\Delta
\tau=7.18$ fm/c started, $\tau_2$ shortly (1 fm/c) after the
finished change, and $\tau_3$ long ($\approx 75$ fm/c) after the
change).
We combine the plots of this propagator with an exponential
$\frac{2}{\sqrt{4m^2-\Gamma^2}}e^{-\frac{1}{2}\Gamma t}$ to show
that it decays with the analytically found decay constant. One can
also nicely see how the wavelength, related to the mass by
$\lambda=\frac{4\pi}{\sqrt{4m^2-\Gamma^2}}$ (see Eq.
(\ref{dglsol})), changes with time and that at earlier times in
the second plot ($\tau_2-t$ large) the frequency still corresponds
to masses close to the initial mass and at times close to the
present to the final mass. This is a nice illustration of the
memory of the retarded propagator and hence the spectral function
in the two-time representation. The stronger retardation ($c>2$)
for larger changes in the self energy can be explained by
considering that the change of the width or mass introduces
additional time scales that might as well become of importance
such that it is no longer so easy to extract the time scale
directly and only from the width.

For the occupation number of the vector mesons, which is basically
given by its propagator $G^<(\tau,\omega)$ and has to be compared
to
$G^<_{\text{Markov}}(\tau,\omega)=2in_{\text{B}}(T(\tau),\omega)A_{\text{Markov}}(\tau,\omega)$
at time $\tau_{\text{off}}$, we find a by about 50\% faster
adaption to medium changes: The quantity is not retarded and gains
information from the future when Fourier transforming. On the
other hand, for the causal dilepton rate (\ref{photrate1}), which
is compared to Eq. (\ref{markovrate}) with the Bose distribution
$n_{\text{B}}(T(\tau),\omega)$ and the spectral function
$A_{\text{Markov}}(\tau,\omega)$ at time $\tau_{\text{off}}$, this
is not the case: We find a retardation much alike that of the
spectral function with similar time scales as given in Eq.
(\ref{timescale}).

\subsection{Quantum interference}
\label{quanint} \label{interference} 
\begin{figure}[htb]
      \begin{center}
        \includegraphics[height=6.5cm]{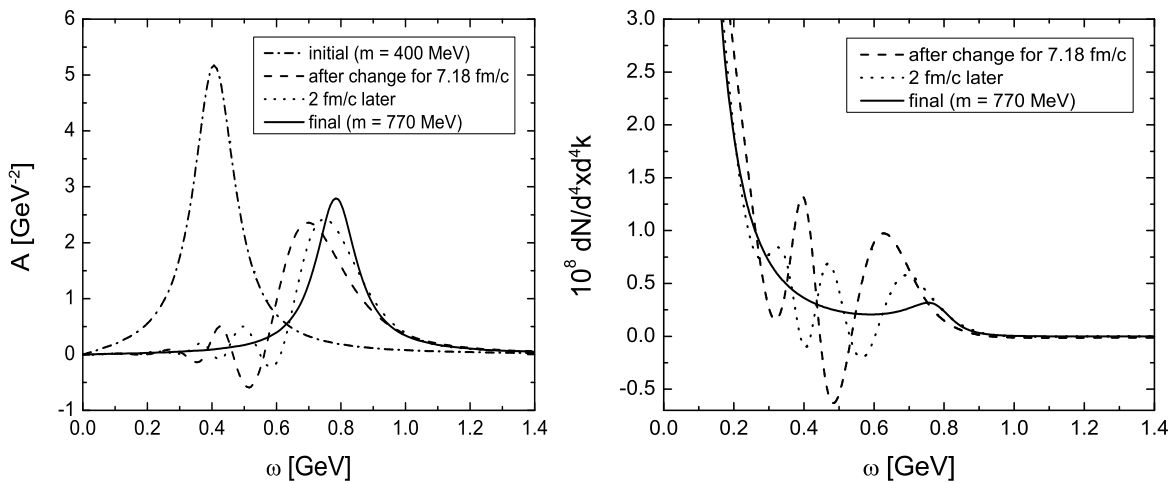}
        \caption{Spectral function and rate for the change of the mass from $m=400$ MeV to 770 MeV (constant $\Gamma$=150 MeV and constant $T=175$ MeV) directly after the self energy has reached its
                 final form (after 7.18 fm/c) and 2 fm/c later. Oscillations and negative values appear in the intermediate spectral functions and rates.}
        \label{fig:oscillations}
      \end{center}
    \end{figure}

\begin{figure}[htb]
      \begin{center}
        \includegraphics[height=6.5cm]{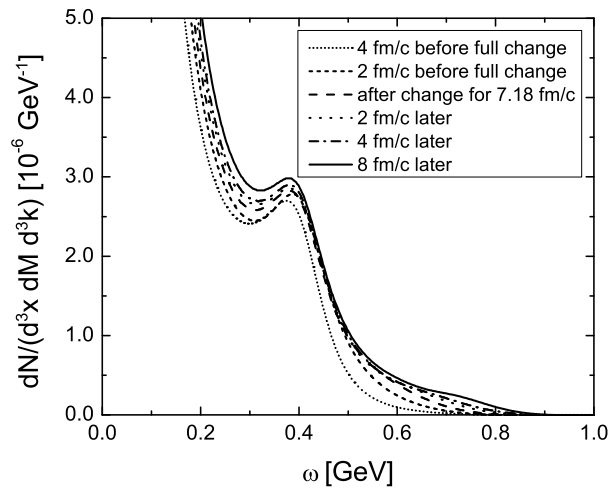}
        \caption{Yields corresponding to rates from Fig. \ref{fig:oscillations} (and some more) integrated over time intervals starting with beginning changes and ending after the
        indicated duration (3.18 fm/c, 5.18 fm/c, 7.18 fm/c, 9.18 fm/c, 11.18 fm/c and 15.18 fm/c).}
        \label{fig:oscillationsyield}
      \end{center}
    \end{figure}

As already mentioned, there
occurs another interesting effect, arising from the quantum
mechanical nature of nonequilibrium dilepton emission.
Oscillations in the changing spectral functions, occupation
numbers and production rates appear as well as interferences that
cannot be present in an approximate semi-classical calculation.
The fact that negative values occur temporarily is comparable to
the well known observation that the Wigner function, the quantum
mechanical analogue to the classical phase space distribution, is
not necessarily positive definite (see e.g.
\cite{Carruthers:1982fa}). It is interesting to note that although
the rates may oscillate below zero, the total yield will stay
positive, being the only physical observable calculated. It can be
shown that the accumulated yield is indeed proportional to the
square of an amplitude \cite{Serreau:2003wr}. The rate in this
case is not to be confused with the rate that results from
measuring the yield and simply dividing by the time interval.
Instead, the rate (\ref{photrate1}) calculated here has the full
quantum mechanical information incorporated and contains
interferences that can cause cancellations - in fact, the rate has
to be able to become negative. An example for the occurring
oscillations is depicted in Fig. \ref{fig:oscillations} for the
spectral function and the rate for the case of a mass shift to 400
MeV in the medium at a constant temperature of $175$ MeV. The
resulting yield integrated over the time interval beginning with
the onset of the changes and ending at the indicated time is shown
in Fig. \ref{fig:oscillationsyield}. It can be nicely seen that
the yield stays positive throughout the evolution while the
temporary rate can become negative.


\subsection{Yields at constant temperature}
\label{sec:consttemp}
\begin{figure}[htb]
      \begin{center}
        \includegraphics[height=12cm]{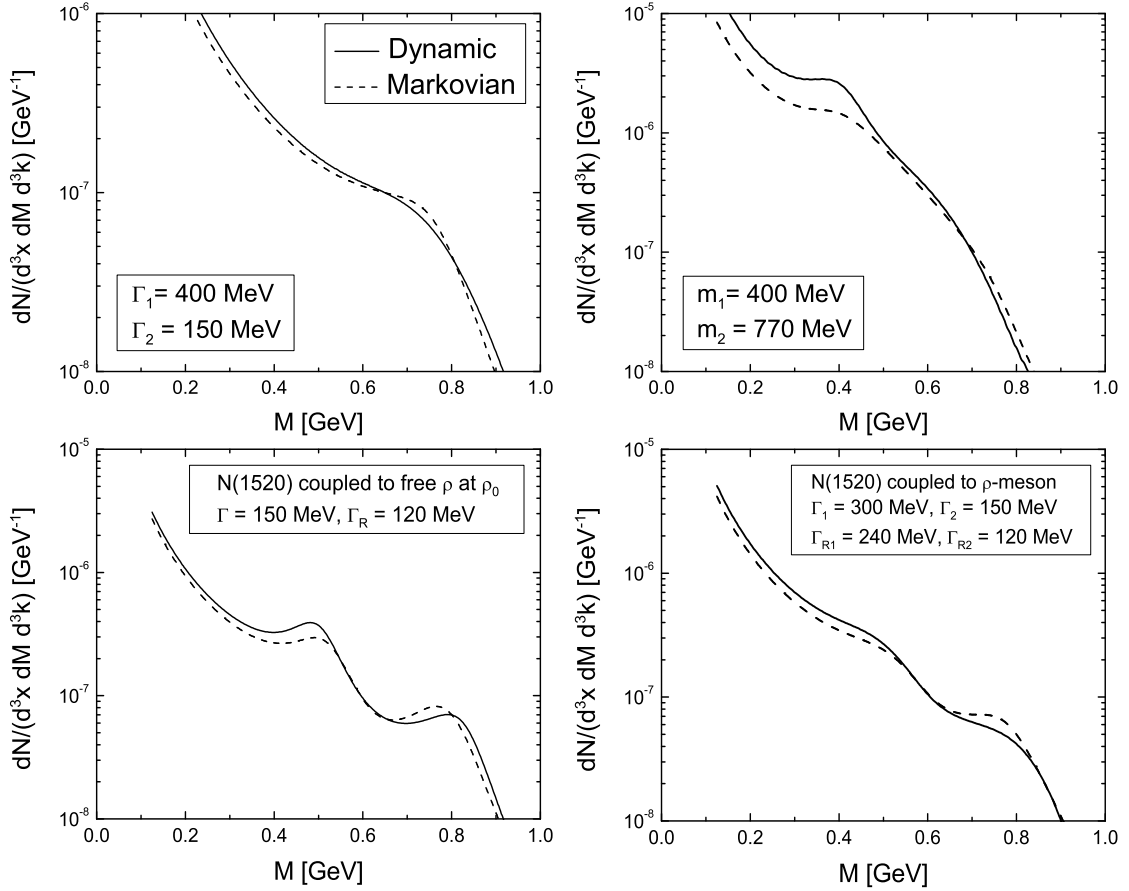}
        \caption{Comparison of the dynamically calculated (solid) to the Markovian (dashed) dilepton yields from an interval of
        duration $\Delta\tau=7.18$ fm/c in that the self energy was changed linearly from initial to final values given in the figure.
        Index 1 indicates initial values, index 2 final ones. The temperature was kept constant at $T=175$ MeV.}
        \label{fig:constTV}
      \end{center}
    \end{figure}

In order to get a first impression of how memory effects can
affect dilepton yields, we investigate yields from systems with a
constant size at constant temperature. This helps to
understand the more complex situation of an evolving fireball,
treated in the next section. For five different scenarios we perform comparisons of the dynamic
to the Markovian calculations in that the spectral properties
adjust instantaneously to the medium and the dilepton rate has no
memory: Modification of the
$\rho$-width to $400$ MeV in the medium, shift of the $\rho$-mass
using a constant coupling at the $\rho$-$\gamma^*$-vertex to $400$
MeV in-medium mass, coupling of the $\rho$ to the N(1520)
resonance with and without additional broadening of the resonance
and the $\rho$-meson. For the $\omega$-meson we consider a mass-shift by 100 MeV to 682 MeV and
broadening to 40 MeV \cite{kl97}. We note that the situation for the $\phi$-meson might be similar.
In each case the vacuum spectral function is approached linearly within the
interval of $7.18$ fm/c, over that we integrate the rates. The
temperature is set to $175$ MeV. The results for the first four
($\rho$-meson-) scenarios are shown in Fig. \ref{fig:constTV}. The
largest, most noticable difference is found for the mass shift of the $\rho$-meson:
The yield from the dynamic calculation is increased by a factor of about 1.8 in the range from
200 to 450 MeV due to the inherent memory integrals. This can be understood since
as opposed to the Markov case the spectral function has a certain memory of the in-medium
conditions and the stronger weighting of lower masses due to the Bose factor increases
the effect additionally.

The other cases show differences, but not as pronounced. We see a
minor enhancement of the lower mass tail due to the Bose factor in
all cases. This is because for the dynamic case there remains a
memory to the larger in-medium width and the resonance peak,
giving more strength to the more strongly weighted lower mass part
of the distribution than in the Markovian case. In the case of
strongly broadened $\rho$-mesons, one can nicely see that the
dynamically calculated yield possesses a broader distribution and
that for the case of the coupling to the N(1520) without
broadening the resonance peak is stronger (by about a factor of
1.5), due to the system's memory to the coupling in the medium.
Here, in the dynamic calculation the vacuum peak is also shifted
further to the right due to memory of the level repulsion effect.
For the case of coupling to the N(1520) with additional broadening
this effect is harder to see because of the very broad
distribution - however, the yield around the resonance peak is
stronger in this case as well.

\begin{figure}[htb]
      \begin{center}
        \includegraphics[height=6.5cm]{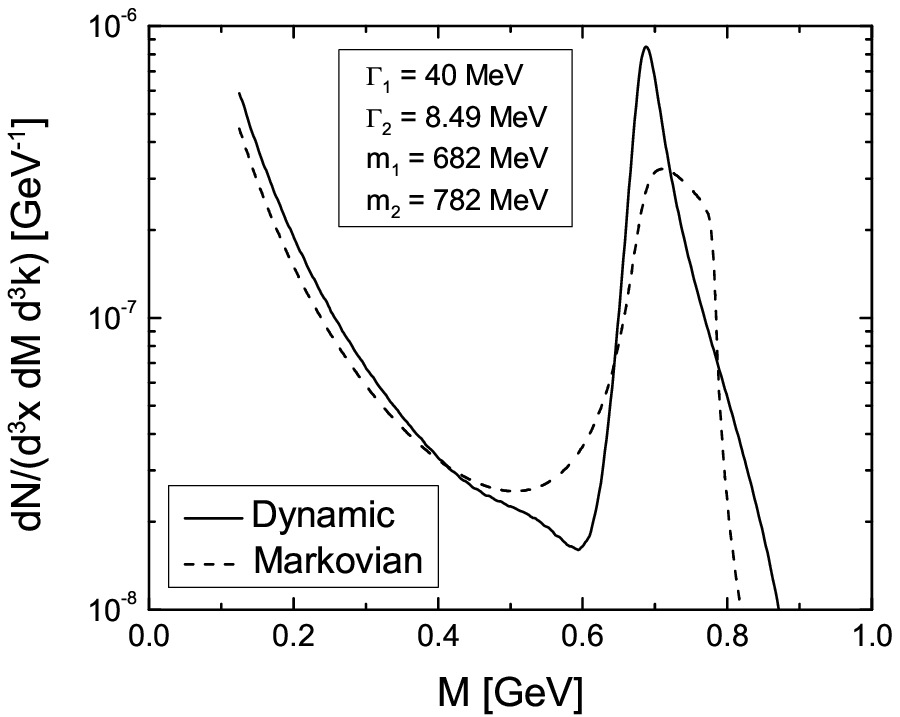}
        \caption{Comparison of the dynamically calculated (solid) to the Markovian (dashed) dilepton yield from $\omega$-mesons
        from an interval of duration $\Delta\tau=7.18$ fm/c in that the self energy was changed linearly from initial to final values given in the figure.
        Index 1 indicates initial values, index 2 final ones. The temperature was kept constant at $T=175$ MeV.}
        \label{fig:constTVomega}
      \end{center}
    \end{figure}

For the $\omega$-meson (Fig. \ref{fig:constTVomega}) large
differences show up due to the strong retardation related to the
small widths involved. One nicely sees that the in-medium peak is
more pronounced in the yield from the dynamic calculation with
memory, whereas the final peak is not yet visible. The stronger
yield above 800 MeV can be explained by the memory to the larger
in-medium width, whereas the decreased yield between 450 and 650
MeV must be due to interferences, discussed in Section
\ref{interference}. The Bose factor does not have that much
influence in this case because the rather small mass shift and
broadening do not produce a spectral function with much weight at
small frequencies.

To conclude this section we state that for the $\rho$-meson we indeed found noticeable modifications of the yields that one would expect qualitatively when including
finite memory for the spectral properties. For the $\omega$-meson, the much longer time scales for the adjustment to the medium further cause structures
resulting from quantum mechanical interference after a comparably quick change over 7.18 fm/c.

\subsection{Fireball model and dilepton yields}
\label{yields}
After having had a first exploration of the influence of finite memory on the dilepton production for constant temperature,
we now study a more realistic situation in that memory effects are expected to be of importance.
The hadronic phase of a fireball, created in a heavy ion collision, lives for about $\tau_{\text{fireball}}=$ 5 to 10 fm/c at SpS energies.
This is the time in that the changes of the vector mesons' self energy take place, a time comparable to the derived
time scale of retardation for the spectral properties and hence the dilepton rate (see Section \ref{sec:adaption}):
\begin{equation}
    \tau_{\text{fireball}}\simeq\bar{\tau}
\end{equation}
This is why the consideration of memory effects is important for heavy ion collisions.

We will model the fireball evolution and convolute it with the calculated time dependent rates.
We choose for the effective volume a longitudinal Bjorken expansion combined with an
accelerating radial flow
\begin{equation}
    V_{\text{eff}}(\tau\geq \tau_0)=\pi c \tau (r_0+v_0(\tau-\tau_0)+0.5 a_0 (\tau-\tau_0)^2)^2\text{,}
    \label{veff}
\end{equation}
with $r_0=6.5$ fm, $v_0=0.15~c$ and $a_0=0.05~c^2/\text{fm}$ (see
also \cite{Greiner:2002nc,Greiner:2001uh}). From (\ref{veff}) and
the constraint of conserved entropy (given by a constant entropy
per baryon $S/A=26$ for SPS energies
\cite{Greiner:1988pc,Cleymans:2001at,Spieles:1997tf}), temperature
and chemical potentials follow as functions of time. The initial
temperature of the fireball is taken slightly above that at
chemical freezeout, 175 MeV, whereas the final temperature,
reached after a lifetime of about 7.18 fm/c is 120 MeV (thermal
freezeout). At this point, we turn off further dilepton
production, by decreasing (quasi instantaneously) the temperature
towards zero. Afterwards only existing vector mesons decay. We
take $A_{\text{B}}=70$ as the number of participant baryons per
unit rapidity and include the 56 lightest baryonic and mesonic
states in this calculation. Using the calculated time dependent
temperature $T(\tau)$ in the computation of the rate and the given
volume $V_{\text{eff}}(\tau)$, we can integrate the rate and get
the accumulated yield per unit four momentum.

\begin{figure}[htb]
  \hfill
  \begin{minipage}[t]{.45\textwidth}
        \includegraphics[height=6.5cm]{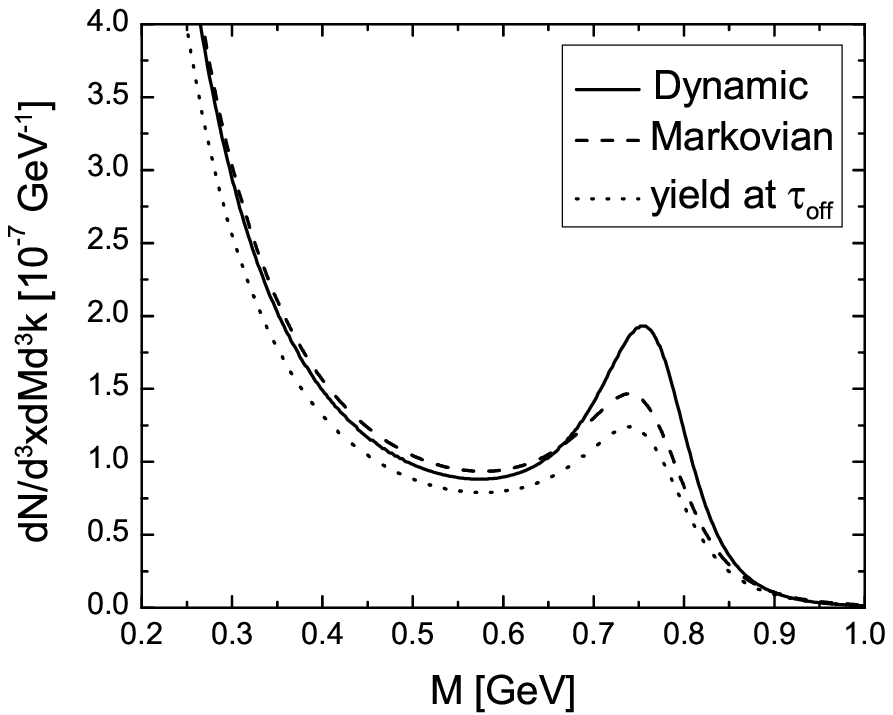}
        \caption{For an equilibrium situation ($\rho$-meson vacuum spectral function and constant temperature) after turning down the temperature, the dynamically calculated yield (solid line)
                 differs from the analytically calculated one (Eq. (\ref{markovdecay})) (dashed line).
                 Also shown is the yield at time $\tau_{\text{off}}$ from an interval starting 7.18 fm/c before (same for both calculations) (dotted line).}
        \label{fig:yieldoff}
  \end{minipage}
  \hfill
  \begin{minipage}[t]{.45\textwidth}
        \includegraphics[height=6.5cm]{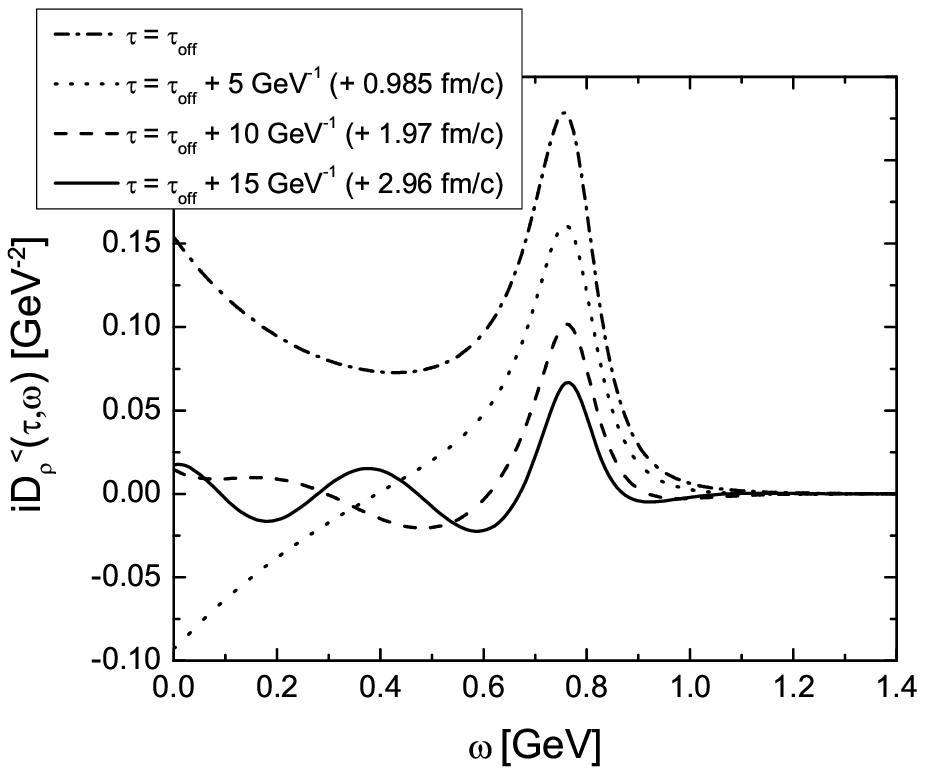}
        \caption{Evolution of $D_{\rho}^{<}(\tau,\omega)$ after having turned off the production at time $\tau_{\text{off}}$.}
        \label{fig:dlessoff}
  \end{minipage}
  \hfill
\end{figure}

Before this we want to consider the dynamic process of decay,
occuring after the production is turned off. Therefore we prepare
a situation of constant temperature and fixed spectral properties
for the $\rho$-meson and turn down the temperature at a certain
point. We then compare the resulting final yield to that
calculated analytically for the Markov case using
\begin{align}
    \frac{dN}{d^3x d^4k}(\tau\rightarrow \infty,\omega)=\frac{dN}{d^3x d^4k}(\tau_{\text{off}},\omega)+
    \left(\frac{1}{\Gamma_{\text{final}}}\right)\frac{2 e^4}{(2\pi)^5}\frac{m_{\rho}^4}{g_{\rho}^2}\frac{1}{\omega^2}n_{\text{B}}(T,\omega)
    \pi A_{\text{Markov}}(\omega)\text{.}
    \label{markovdecay}
\end{align}
The result is given in Fig. \ref{fig:yieldoff}. It nicely shows
the difference in the decay process: The dynamic calculation leads
to a much more pronounced peak while the Markov approximation has
larger contributions away from the peak. This can be understood by
the investigating the dynamic evolution of the occupation of
$\rho$-mesons, given by $D_{\rho}^{<}(\tau,\omega)$, after turning
down the temperature. We show this in Fig. \ref{fig:dlessoff}. It
nicely exhibits how the "thermal" occupation far from the pole
mass disappears. This is not given in the approximate Markov
formula (Eq. (\ref{markovdecay})), where $n_{\text{B}}(\omega)$ is
constant in time. Integrating the yield over invariant masses $M$
from 125 to 1500 MeV yields the same overall dilepton number for
both cases within a percent. Furthermore we find the expected
behavior of the dilepton yield integrated over the invariant mass
after the time of switching off $\tau_{\text{off}}$:
Its increase per unit time decreases with $e^{-\Gamma t}$ (for $t>\tau_{\text{off}}$).

\begin{figure}[htb]
      \begin{center}
        \includegraphics[height=12cm]{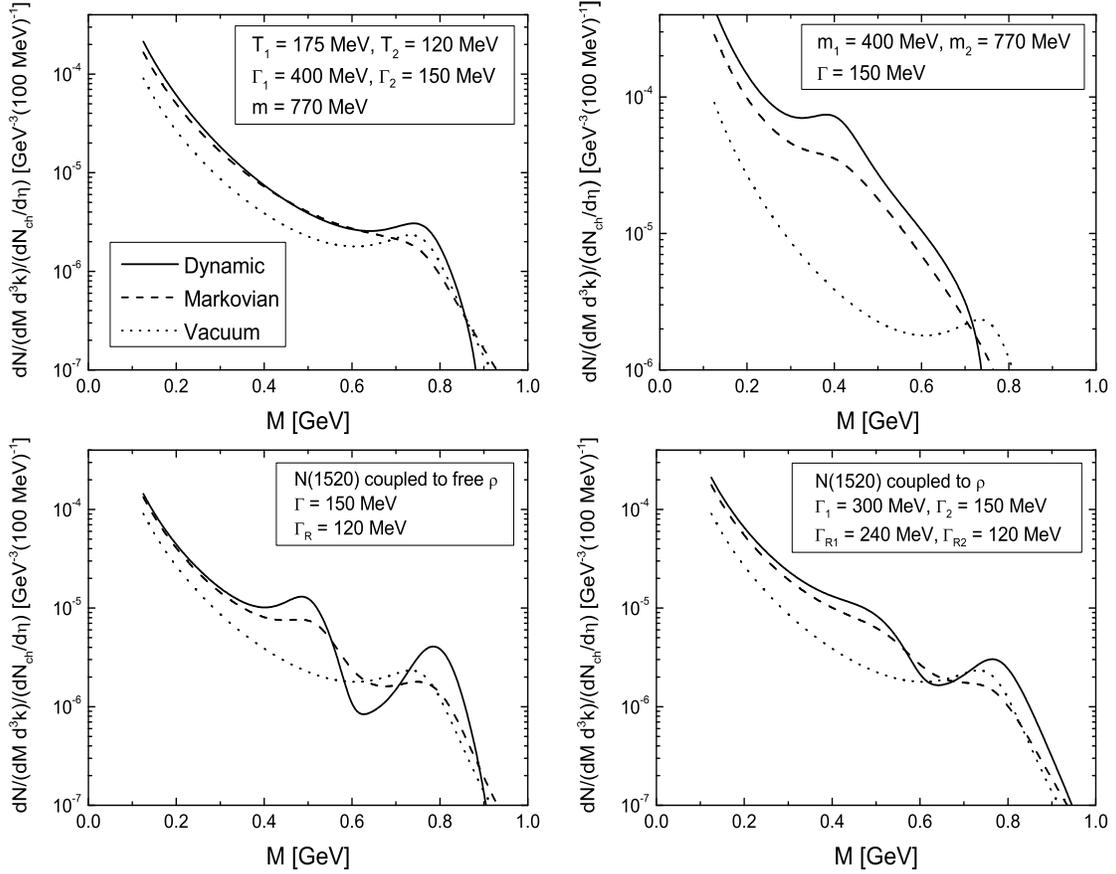}
        \caption{Comparison of the dynamically calculated (solid) to the Markovian (dashed) final dilepton yields where
        the self energy was changed linearly over an interval of duration $\Delta\tau=7.18$ fm/c as indicated in the corresponding figure (index 1 for initial, 2 for final quantities).
        The time dependent rate was then convoluted with the temperature and volume of the fireball. Dotted lines show yields from unmodified (vacuum) $\rho$-mesons.}
        \label{fig:fireball}
      \end{center}
    \end{figure}

Now we turn to the situation of an expanding fireball.
We consider five different scenarios as in the constant temperature and volume case (see Section \ref{sec:consttemp}): Broadening of the $\rho$-meson to a width of
400 MeV, a mass-shift of the $\rho$-meson to 400 MeV mass, coupling of the $\rho$-meson to the N(1520) resonance
with and without broadening, and modification of mass and width of the $\omega$-meson. Comparison to calculations at constant temperature reveals the
additional influence of dropping temperature and increasing volume, as well as that of the mentioned difference in behavior to the Markov calculation
after the production has been turned off.
Overall the differences in the yields are stronger for the case of changing temperature and volume. Fig. \ref{fig:fireball} shows the comparison
to the Markov calculation for the four $\rho$-meson scenarios.
There always remains a certain memory of the higher temperatures at earlier times which together with the increasing volume leads to an
increased yield within the intermediate mass regime from 400 to 700 MeV for the case of a mass shift to 400 MeV in-medium mass, which was mostly restricted to the vicinity of
400 MeV in the constant temperature case. The dynamically calculated yield around the in-medium peak is now a factor of two larger than that from the
Markovian calculation. This constitutes a significant effect and shows the importance of the consideration of the full dynamics for mass shifts,
especially if one asks for precise theoretical predictions.
In the case of coupling to the N(1520) resonance with and without broadening one now gets more pronounced s-shape structures in the yield than in
the Markov case. In addition the different behavior after the temperature was turned off becomes visible in the peak at the vacuum mass.
This effect even dominates for the case of the broadened $\rho$-meson such that the actual effect of a broader distribution in the dynamic case, which was seen
in the constant temperature calculation, is hidden.

\begin{figure}[htb]
      \begin{center}
        \includegraphics[height=7cm]{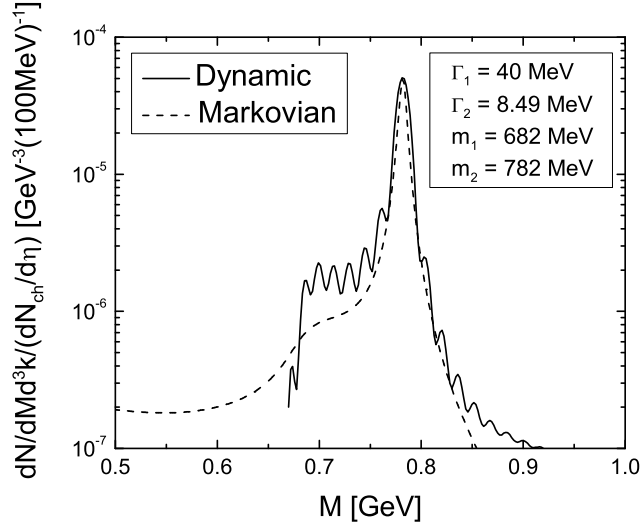}
        \caption{Comparison of the dynamically calculated (solid) to the Markovian (dashed) final dilepton yields for the $\omega$-meson.
        The self energy was changed linearly over an interval of duration $\Delta\tau=7.18$ fm/c as indicated in the figure (index 1 for initial, 2 for final quantities).
        The time dependent rate was then convoluted with the temperature and volume of the fireball.}
        \label{fig:fireballomega}
      \end{center}
    \end{figure}

For the situation of a propagating $\omega$-meson mode (Fig.
\ref{fig:fireballomega}) we calculate the final Markovian yield by
adding the contribution given in Eq. (\ref{markovdecay}) but using
the initial temperature of 175 MeV, since most of the states,
populated at high temperature, are still decaying at the time of
thermal freeze-out and contribute to the final yield. One can see
an enhancement by a factor of more than two between the in-medium
mass and about 750 MeV in the dynamic calculation, caused by the
finite memory to the in-medium properties. Due to the finite
resolution, the numerical treatment can only approximate the
narrow $\omega$ vacuum peak and gives visible wiggles in the
yield.

What about the non-zero momentum modes? The shown calculations for the momentum mode $\textbf{k}=0$ can be extrapolated at least to further lower momentum modes, since the self energy is a continuous
function with respect to momentum. Typically, for the lowest momenta the strongest medium modifications are seen in experiment \cite{Agakishiev:1997au,Agakishiev:2005nm}.
The yield per unit rapidity and invariant mass can then be given by converting variables and approximating the integral over transverse
momenta by a product of the yield for momentum $\textbf{k}=0$ with the regarded momentum range $\Delta \textbf{k}_{\perp}$.
In order to get the yield for $y=0$ in the center of mass frame, in the Bjorken model one has to integrate over all rapidities of the fluid cells, using that $y=\eta+y'$,
where $\eta$ is the space-time rapidity of the cell and $y'$ the momentum rapidity of the particles in the restframe of the cell.
\begin{align}
 \left.\frac{dN}{dM dy}\right|_{k_{\perp}<k_{\text{cut}},y=0,\text{CM}}&=M\int_{-\infty}^{\infty}d\eta\int_{-\infty}^{\infty}dy'
 \int_0^{k_{\text{cut}}}d^2k_{\perp}
 \left(\frac{dN}{dE d^3 k}\right)_{\text{LR}, k_{\parallel}(y'),k_{\perp}} \delta(\eta+y')\notag\\
 &\approx M \int_{-\infty}^{\infty}d\eta\int_{-\infty}^{\infty}dy'
 \left(\frac{dN}{dE d^3 k}\right)_{\text{LR}, k_{\parallel}(y'),k_{\perp}=0} \delta(\eta+y')\pi k_{\text{cut}}^2
\end{align}
where CM indicates the center of mass frame, LR the local restframe and $k_{\text{cut}}$ is the upper momentum cutoff for transverse momenta.
The cutoff $k_{\text{cut}}$ is chosen to be 500 MeV in \cite{Agakishiev:1997au,Agakishiev:2005nm} for investigation of dilepton yields for lower transverse momenta.
For the Bjorken model we encounter the problem that arbitrarily large rapidities $y'$ and hence also large longitudinal momenta contribute.
Approximating the yields for these momentum modes by that for $\textbf{k}=0$ is not reasonable.

On the other hand for a "semi-static" expanding fireball model as employed in e.g. \cite{Rapp:1999ej,Rapp:1999us} the situation becomes much simpler
as in this case the only dilepton longitudinal momentum contributing to the midrapidity yield is $k_{\parallel}=0$
and we have
\begin{align}
 \left.\frac{dN}{dM dy}\right|_{k_{\perp}<k_{\text{cut}},y=0}&=M\int_0^{k_{\text{cut}}}d^2k_{\perp}\left(\frac{dN}{dE d^3 k}\right)_{k_{\parallel}=0,k_{\perp}}\notag\\
 &\approx M\left(\frac{dN}{dE d^3 k}\right)_{k_{\parallel}=0,k_{\perp}=0}\pi k^2_{\text{cut}}\text{.}
\end{align}
No momenta larger than $k_{\text{cut}}$ are involved for this case and one can approximate the yields for the appearing small momenta by those for zero momentum.
Hence also for measurable
finite momentum intervals the resulting modifications of the yields due to finite memory are expected to be completely similar and our primary message remains valid.

We summarize by stating that inclusion of the fireball evolution leads to further enhancement of the differences between the dilepton yields calculated
using a full nonequilibrium quantum field theoretical formulation including retardation and its much simpler Markovian approximation.
For the realistic situation of a heavy ion collision memory effects have to be considered when it comes to predicting medium
modifications of particles.



\section{Summary and Conclusions}
\label{conclusion}
    In the present work, we have numerically calculated dilepton
    production rates within a nonequilibrium field theory formalism,
    based on the real time approach of Schwinger and Keldysh \cite{Schw61,Ke64,Ke65,Cr68}.
    We employed the Kadanoff-Baym equations, generalized to the relativistic Dirac structure,
    to derive the formula for the dilepton production rate, which is non-local in time
    and hence includes the usually disregarded finite memory. The rate is in principle
    a (half) Fourier transform over past times of the virtual photon occupation, described by the Green function $D_{\gamma}^<$ in the two-time representation.
    Medium modifications of the vector mesons enter this
    production rate via the virtual photon's self energy that is connected to the vector meson propagator by the principle of vector meson dominance.

    The equation of motion for the retarded and advanced vector meson propagator also followed from the Kadanoff-Baym equations.
    From its solution it was possible to extract the dynamic behavior of the meson's spectral function and to define a time scale on that it adjusts to medium
    changes, which were incorporated by the definition of a time dependent self energy for the vector mesons.
    This time scale was found to be proportional to the inverse vacuum width of the
    meson like $c/\Gamma$. $c$ lies between $2$ and about $3.5$.
    The time that the dilepton rate needs to follow changes was found to be approximately equal to that for the spectral function.
    Since these time scales lie in the range of typical times in that the hadronic phase of fireballs
    in heavy ion reactions exists and evolves, we expected an influence of the found retardation on the dilepton yields and investigated different scenarios
    quantitatively.
    Possible medium modifications of the vector mesons, as shifted pole masses, motivated by Brown-Rho scaling,
    broadening and coupling to resonance-hole pairs were considered.
    Investigation of the dynamical off-shell evolution of the spectral function, meson occupation number and resulting dilepton rate
    revealed the quantum mechanical nature of the system, manifested in appearing oscillations and
    hence interferences in the regarded quantities after changes of the self energy that were fast as compared to the derived time scales.
    All quantities were found to possibly become negative, which is a major difference to the Markov approximation
    within which all quantities are always positive definite.
    The oscillations potentially cancel when the rate is integrated over time such that the measurable
    dilepton yield is found to be positive, as it has to be.
    On the other hand, for situations where the particular rate can become negative, a semi-classical interpretation with positive definite rates,
    as employed in present day transport codes, is not possible.

    We first calculated dilepton yields from a time dependent system at constant volume and temperature considering different possible medium modifications
    and compared to the yields found assuming an instantaneous (Markovian) adjustment of all quantities to the medium. For the $\rho$-meson we found
    qualitatively expected results for each case. Quantitatively, the strongest difference in the differently calculated yields
    appeared for the case of a mass shift to 400 MeV in-medium mass, as suggested by Brown-Rho scaling.
    This is due to the stronger weighting of the lower masses by the Bose factor and since the
    spectral function moves more slowly towards the free mass in the dynamic situation, we found more enhancement in the yield than for the Markov calculation
    without memory. It was shown that there can be more than a factor of two difference. For the medium modified $\omega$-meson the qualitatively expected results
    plus additional structures caused by quantum interferences were seen.
    These are stronger than in the case of the $\rho$-meson because the performed change is faster relative to the time scale for adiabatic
    behavior of the $\omega$-meson, which has a very small width.

    The introduction of a fireball model allowed us to perform more realistic simulations including changing temperature and volume,
    adjusted to the case of the SpS with energies of 158 AGeV.
    The modifications of the yields due to inclusion of finite memory were enhanced as compared to the constant temperature case. Also the behavior
    of the regarded quantities after freeze-out was revealed to be different from that assumed in Markovian calculations.
    For the mass shift of the $\rho$-meson a factor of two difference in the two calculated yields was found around the in-medium mass - a significant effect.
    For the case of substantial broadening without further modifications, the differences are only very small.
    In all scenarios, also for the $\omega$-meson, lower invariant masses were slightly enhanced in the dynamic calculation.
    For the coupling to the N(1520) resonance the peaks were more pronounced than in the Markov calculation.

    In summary, our findings show that exact treatment of medium modifications of particles in relativistic heavy ion collisions requires
    the consideration of memory effects. This is in particular true for mass shifts or the occurrence of two-peak structures
    in their spectral properties.
    It has to be a future task to find how to incorporate such memory effects in semi-classical transport simulations. This should be also of relevance for
    the description of production and propagation of vector mesons through cold nuclei in photo-nuclear reactions \cite{Muhlich:2002tu,Muhlich:2003tj}.

\subsection*{Acknowledgments}
The authors thank Stefan Leupold and Marcus Post for helpful discussions throughout this work.

\appendix

\section{On the acausality of the Kadanoff-Baym parametrization}
\label{app:acausal}
    In this appendix we first show that the choice of time variables $\tau_{\text{KB}}=\frac{t_{\alpha}+t_{\beta}}{2}$ and $\Delta t=t_{\alpha}-t_{\beta}$ leads
    to an acausal spectral function $A(\tau_{\text{KB}},\omega)$. This choice will be referred to as the Kadanoff-Baym parametrization
    since it was first introduced by Kadanoff and Baym in \cite{kb62} for a first order gradient expansion of the full quantum transport equations.
    We introduce a different parametrization $\tau=t_{\alpha}$ and $\Delta t=t_{\alpha}-t_{\beta}$
    and obtain a spectral function $A(\tau,\omega)$ that has no information on the future incorporated.

    The following analysis of the differential equation of the retarded propagator
    \begin{equation}
        \hat{D}_{t_A} D_{\rho}^{\text{ret}}(t_A,t_B)=\int_{t_B}^{t_A}dt_2\Sigma_{\rho}^{\text{ret}}(t_A,t_2)D_{\rho}^{\text{ret}}(t_2,t_B)
    \end{equation}
    allows us to find the times, from which medium information $\Sigma^{\text{ret}}(t_1,t_2)$ is contributed to $D_{\rho}^{\text{ret}}(t_A,t_B)$.

\begin{figure}[H]
  \hfill
  \begin{minipage}[t]{.45\textwidth}
      \includegraphics[height=6.5cm]{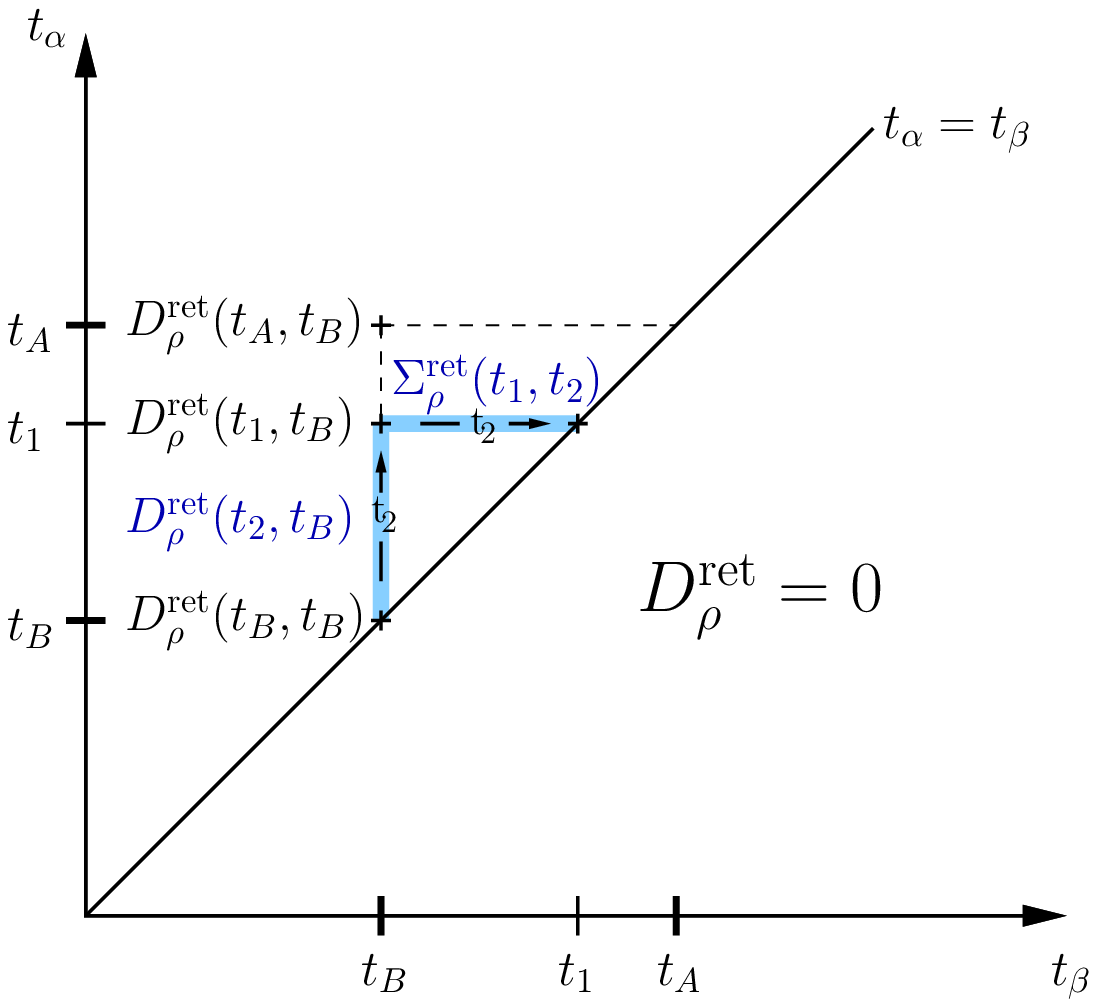}
      \caption{(Color online) Ranges in time (shaded regions) that contribute to $D_{\rho}^{\text{ret}}(t_1,t_B)$ in the $t_{\alpha}$-$t_{\beta}$-plane,
    where $t_{\alpha}$ ($t_{\beta}$) stands for the first (second) time argument.}
      \label{fig:sketch1}
  \end{minipage}
  \hfill
  \begin{minipage}[t]{.45\textwidth}
      \includegraphics[height=6.5cm]{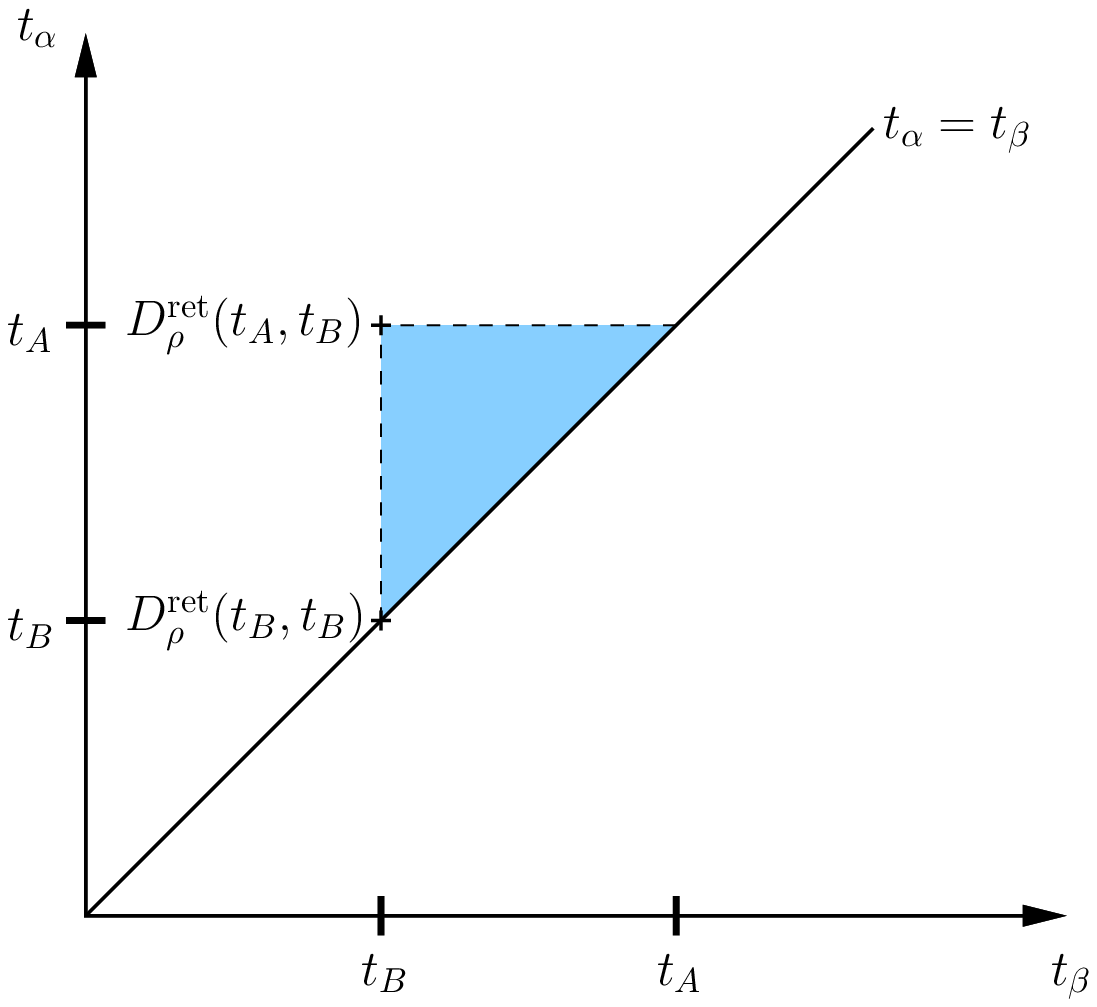}
      \caption{(Color online) Contributions of $\Sigma^{\text{ret}}(t_{\alpha}$, $t_{\beta})$ to $D_{\rho}^{\text{ret}}(t_A,t_B)$ come from times $t_{\alpha}$, $t_{\beta}$, lying within the shaded region.}
      \label{fig:triangle}
  \end{minipage}
  \hfill
\end{figure}

    $D_{\rho}^{\text{ret}}(t_A,t_B)$ follows by integration of
    $
        \hat{D}_{t_1} D_{\rho}^{\text{ret}}(t_1,t_B)=\int_{t_B}^{t_1}dt_2\Sigma_{\rho}^{\text{ret}}(t_1,t_2)D_{\rho}^{\text{ret}}(t_2,t_B)
    $
    over $t_1\epsilon[t_B,t_A]$, starting at $t_1=t_B$\footnotemark.
    \footnotetext{For the case of $\hat{D}_{t_1}=(-\partial^2_{t_1}-m^2)$ the initial conditions are $D_{\rho}^{\text{ret}}(t_B,t_B)=0$ and
    $\partial_{t_1}D_{\rho}^{\text{ret}}(t_1,t_B)|_{t_1=t_B}=-1$.}
    Since all quantities are retarded, we always have $t_A\geq t_1\geq t_2 \geq t_B$.
    $D_{\rho}^{\text{ret}}(t_1,t_B)$ itself has contributions from $D_{\rho}^{\text{ret}}(t_2,t_B)$ and $\Sigma_{\rho}^{\text{ret}}(t_1,t_2)$ with
    $t_2\epsilon[t_B,t_1]$.  Fig. \ref{fig:sketch1} schematically shows where in time these contributions are located.
    On the horizontal line enter contributions from $\Sigma_{\rho}^{\text{ret}}(t_1,t_2)$, on the vertical line
    those from $D_{\rho}^{\text{ret}}(t_2,t_B)$. After integration over
    $t_1\epsilon[t_B,t_A]$, one finds that
    $D_{\rho}^{\text{ret}}(t_A,t_B)$ is created solely with information
    $\Sigma_{\rho}^{\text{ret}}(t_1,t_2)$,
    defined at times lying in the triangle shown in Fig. \ref{fig:triangle}.

\begin{figure}[H]
  \hfill
  \begin{minipage}[t]{.45\textwidth}
      \includegraphics[height=6.5cm]{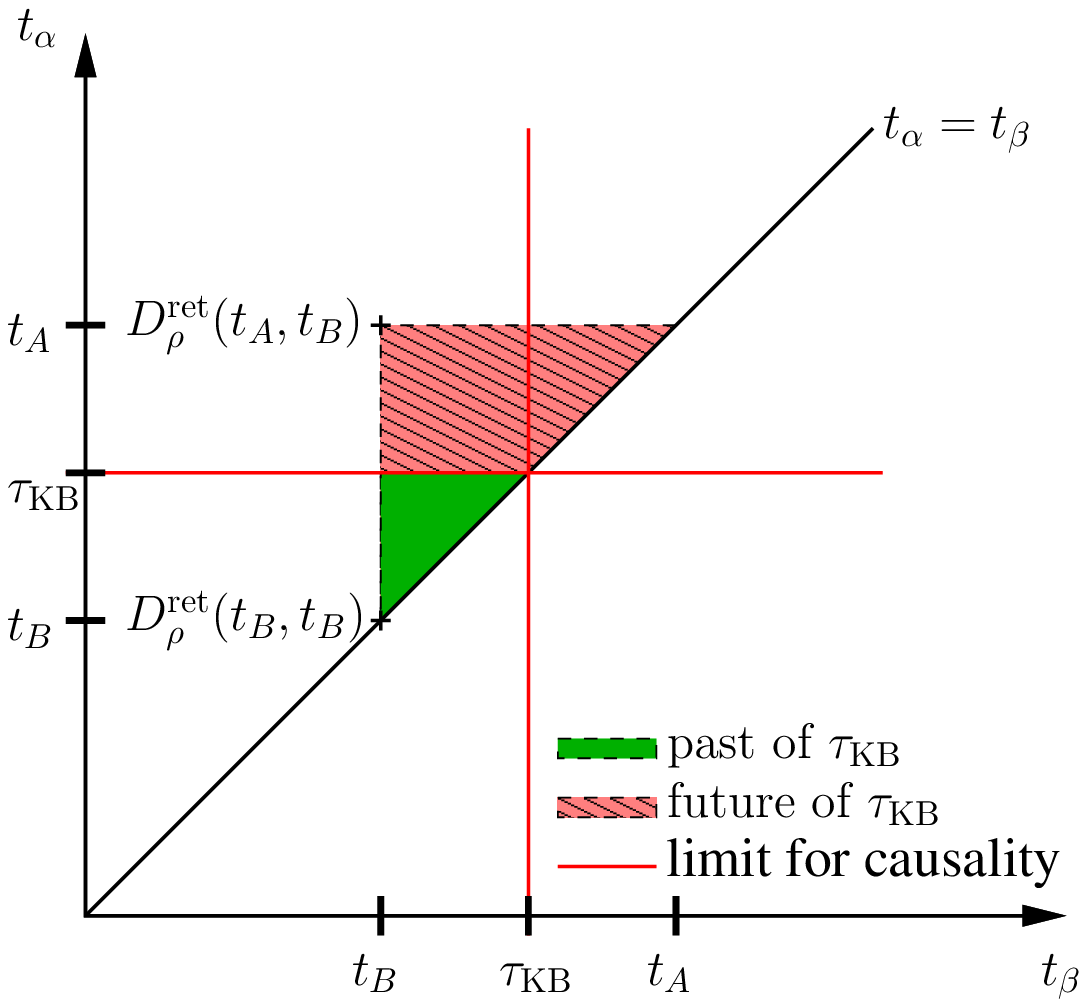}
      \caption{(Color online) After transformation to $\tau_{\text{KB}}=\frac{t_{\alpha}+t_{\beta}}{2}$ and $\Delta t=t_{\alpha}-t_{\beta}$ the Green function at time
      $\tau_{\text{KB}}=\frac{t_{A}+t_{B}}{2}$, $D_{\rho}^{\text{ret}}(\tau_{\text{KB}},\Delta t)$,
      has contributions of $\Sigma^{\text{ret}}(t_{\alpha}$, $t_{\beta})$ also from its future, indicated by the upper shaded region.}
      \label{fig:KBtriangle}
  \end{minipage}
  \hfill
  \begin{minipage}[t]{.45\textwidth}
      \includegraphics[height=6.5cm]{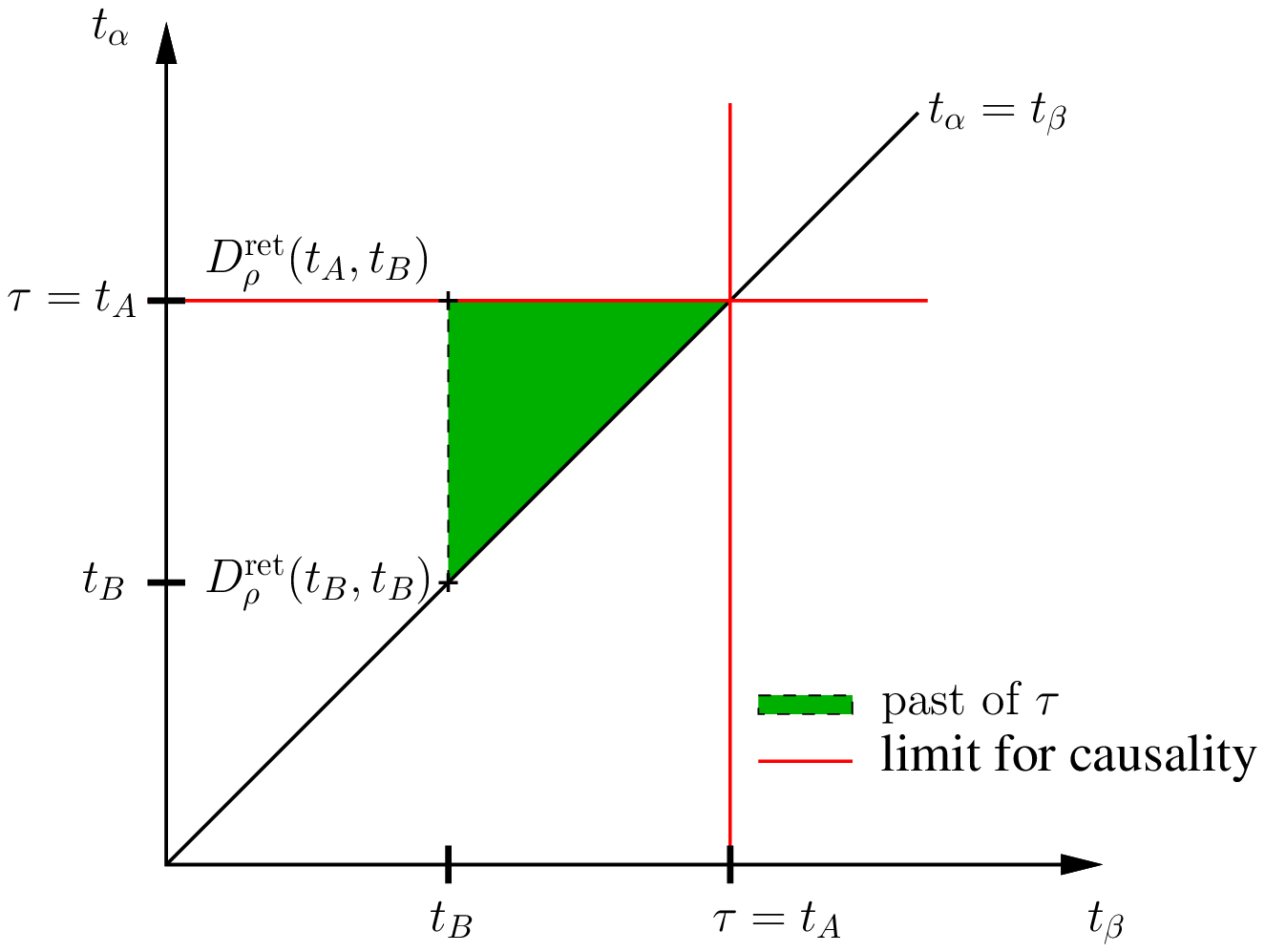}
      \caption{(Color online) Choosing the parametrization $\tau=t_{\alpha}$ and $\Delta t=t_{\alpha}-t_{\beta}$ leads to a Green function $D_{\rho}^{\text{ret}}(\tau=t_A,\Delta t)$
      without contributions of $\Sigma^{\text{ret}}(t_{\alpha}$, $t_{\beta})$ from the future.}
      \label{fig:CGtriangle}
  \end{minipage}
  \hfill
\end{figure}

    When we now choose to transform time variables like $\tau_{\text{KB}}=\frac{t_{\alpha}+t_{\beta}}{2}$ and $\Delta t=t_{\alpha}-t_{\beta}$,
    we find that the Green function $D_{\rho}^{\text{ret}}(\tau_{\text{KB}},\Delta t)$ is created also with information defined in its future as demonstrated in
    Fig. \ref{fig:KBtriangle}. Hence, this parametrization cannot be causal and is not suitable for finding appropriate time scales for adaption of
    the spectral function to medium changes in time. The spectral function $A(\tau_{\text{KB}},\omega)=-\frac{1}{\pi}\text{Im}D_{\rho}^{\text{ret}}(\tau_{\text{KB}},\omega)$
    is the imaginary part of the Fourier transform
    in relative time coordinates of the Green function and hence would have incorporated information on self energies from the future.

    On the other hand, using $\tau=t_{\alpha}$ and $\Delta t=t_{\alpha}-t_{\beta}$, as done throughout this work, the situation
    looks as depicted in Fig. \ref{fig:CGtriangle}. Only past times contribute to the Green function at time $\tau=t_A$.
    A spectral function defined as $A(t_{\alpha},\omega)=-\frac{1}{\pi}\text{Im}D_{\rho}^{\text{ret}}(t_{\alpha},\omega)$ is causal as required.

\bibliography{diplom}

\end{document}